\documentclass[aps,english,superscriptaddress,preprintnumbers,reprint,footinbib,amsmath,amssymb,pra,longbibliography]{revtex4-2}
\usepackage{graphicx}
\usepackage[version=3]{mhchem}
\usepackage{graphicx}
\usepackage{dcolumn}
\usepackage{bm}
\usepackage{hyperref}
\hypersetup{
    colorlinks=true,
    linkcolor=blue,
    filecolor=blue,
    urlcolor=blue,
   citecolor=blue}
\usepackage{enumitem}
\usepackage{color}
\usepackage{amsmath}
\usepackage{tikz-feynman}
\usepackage{comment}

\begin{document}
\title{Schottky anomaly in a cavity-coupled double quantum well}

\author{Valerii K. Kozin}
\affiliation{Department of Physics, University of Basel, Klingelbergstrasse 82, CH-4056 Basel, Switzerland}
\author{Dmitry Miserev}
\affiliation{Department of Physics, University of Basel, Klingelbergstrasse 82, CH-4056 Basel, Switzerland}
\author{Daniel Loss}
\affiliation{Department of Physics, University of Basel, Klingelbergstrasse 82, CH-4056 Basel, Switzerland}
\author{Jelena Klinovaja}
\affiliation{Department of Physics, University of Basel, Klingelbergstrasse 82, CH-4056 Basel, Switzerland}

\date{\today}

\begin{abstract}
We present a theoretical study of a mesoscopic two-dimensional electron gas confined in a double quantum well that is coupled to a uniform quasi-static cavity mode via fluctuations of the dipole moment.  We focus on the regime of large number of electrons participating in the virtual intersubband transitions. In this regime, the effective photonic potential is no longer quadratic but, instead, it contains large number of minima. Each minimum represents a nearly harmonic oscillator with the renormalized cavity frequency that is much greater than its bare value. The energy offset of a minimum scales quadratically with respect to the photon coordinate corresponding to this minimum. These energy offsets determine the statistical weight of each minimum, and altogether they result in the additive correction to the heat capacity of the system. This correction exhibits a Schottky anomaly and a $0.5 k_B$ plateau at low temperatures. This behavior can be associated with the emergence of a new degree of freedom. This degree of freedom does not manifest in the optical conductivity and can only be observed via the heat capacity measurement.
\end{abstract}

\maketitle

\section{Introduction}
 
Cavity-coupled condensed matter systems offer a versatile framework for designing novel states of matter~\cite{cavQuantMatReview,QIN20241}. For instance, resonant cavity driving enables Floquet physics, which encompasses phenomena such as topological insulators~\cite{PhysRevB.81.165433,PhysRevB.79.081406,Lindner2011,Dehghani2015}, topological superconductors~\cite{PhysRevB.95.155407,PhysRevLett.116.176401,PhysRevB.88.155133,PhysRevLett.111.136402,Ricco2022, PhysRevA.106.023702}, and Floquet-engineered topological band structures~\cite{doi:10.1126/science.1239834,McIver2020,PhysRevB.97.035416}. Engineering the properties of matter with optical cavities goes beyond solid-state systems and is used in atomic setups~\cite{RevModPhys.91.015005,Roux2020,PhysRevLett.118.073602,PhysRevLett.125.263606}.

Recent theoretical advances have proposed cavity-modified~\cite{kozin2024cavityenhancedsuperconductivitybandengineering,Sentef2018,PhysRevLett.125.217402,DrivingFluctuationsPairing, DrivingPairing,PhysRevB.104.L140503,eckhardt2023theory,PhysRevLett.104.106402} superconductivity in cavity-coupled two-dimensional electron gases (2DEG), driving growing interest in these systems. The emergence of novel phenomena in cavity-coupled systems frequently relies on the strong light-matter coupling regime~\cite{FriskKockum2019}. This regime can be achieved through cavity pumping~\cite{FloquetReview}, tuning cavities to resonate with plasmons or exciton-polaritons~\cite{MicrocavitiesKavokin,PhysRevB.98.125115}, or by reducing the effective mode volume in specially engineered resonators~\cite{ModeCompressionPRB,ModeCompressionNanoLett}.

Unlike traditional Fabry-Pérot cavities~\cite{MicrocavitiesKavokin}, metallic resonators with compressed mode volumes behave as $LC$ circuits with a frequency $\omega_0 = 1 / \sqrt{LC}$ that is not directly linked to the resonator size. Such resonators enable ultra-strong light-matter coupling, significantly altering electronic systems in their ground states without any driving, such as evidenced in quantum Hall~\cite{FaistHallBreakdown} or anomalous Hall ~\cite{PhysRevB.99.235156} systems. Such an approach resembles the Lamb shift and can manifest itself in inducing quantum phase transitions~\cite{PhysRevResearch.6.033188} due to cavity renormalization of the electronic structure~\cite{PhysRevB.108.085410,Dmytruk2022}.

In this paper, we consider a 2DEG confined in a wide double quantum well (DQW) that is coupled to an $LC$-cavity~\cite{PhysRevX.4.041031,PhysRevResearch.6.033097,PhysRevB.85.045304,PhysRevLett.105.196402,PhysRevLett.102.186402,PhysRevB.91.125409,Goulain2023} - a sub-wavelength resonator allowing for achieving high mode-volume compression~\cite{PoliniAndolinaAmperian2022}.
The coupling is generated via the intersubband transitions of the DQW~\cite{TodorovExper2019}. We study the contribution of the cavity photons and the light-matter interaction to the heat capacity of the whole system.  In the regime of large number of electrons participating in the virtual intersubband transitions, the effective potential that describes the cavity subsystem is no longer harmonic but instead, it contains large number of minima.
Each minimum can be approximated by a harmonic potential with renormalized cavity frequency representing a {polariton}. The finite energy offset of each minimum is proportional to the square of the photon coordinate at the minimum. These minima offsets determine the statistical weight of each near-harmonic minimum which results in an additive contribution to the heat capacity that contains the Schottky anomaly and a $0.5 k_B$ plateau at low temperatures, $k_B$ is the Boltzmann constant.  We attribute this behavior to the emergence of a new degree of freedom that does not couple to the electric current, i.e. it is electrically "neutral".  Therefore, this emergent degree of freedom cannot be detected via the optical conductivity or the absorption spectra, and exclusively appears in the heat capacity measurement. We study this problem beyond the mean-field approximation, taking into account the leading interaction corrections.

\section{Theoretical Model}
\label{sec:model}
\begin{figure}[h]
	\includegraphics[width=0.98\columnwidth]{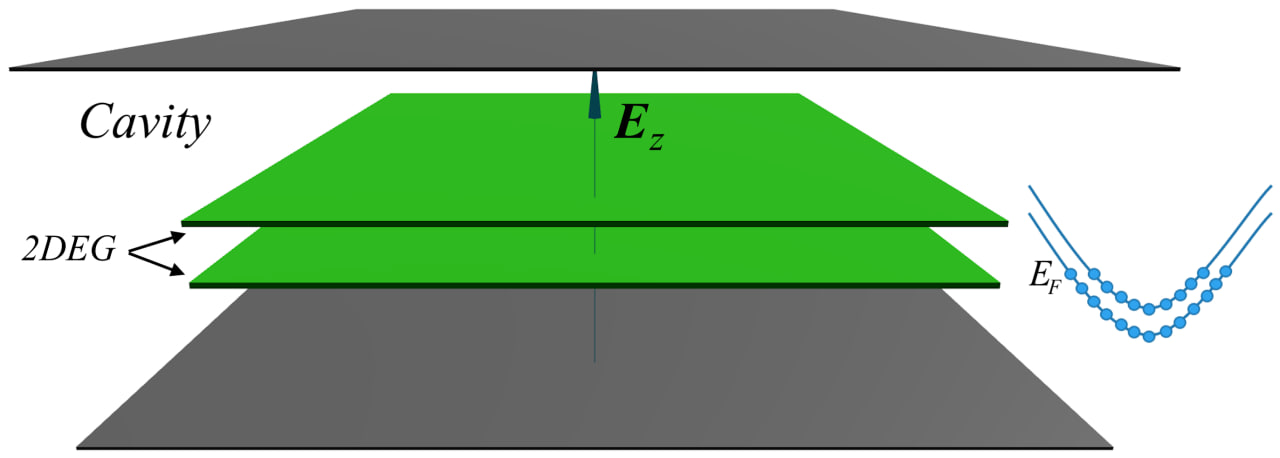}
	\caption{\label{fig:1} 
Sketch of the system: DQW 2DEG is placed parallel to the capacitor plates of the $LC$-cavity. The electric field operator $\hat{E}_z$ (black arrow) of the cavity mode is polarized perpendicular to the DQW 2DEG plane. Inset: two lowest electron subbands are populated by electrons up to the Fermi energy $E_F$.}
	\vspace{-5pt}
\end{figure}

We consider a spin-degenerate 2DEG localized in a wide DQW that is coupled to a uniform LC cavity mode polarized across the DQW, see Fig.~\ref{fig:1}. 
{Similar systems are commonly described in the length (dipole) gauge~\cite{PhysRevX.4.041031,PhysRevResearch.6.033097,PhysRevB.85.045304,PhysRevLett.105.196402,PhysRevLett.102.186402,PhysRevB.91.125409}. For a theoretical treatment of cavity-coupled mesoscopic systems, however, the velocity gauge where the coupling enters via the Peierls substitution is convenient for a semiclassical analysis. We thus start with the following Hamiltonian:}
\begin{eqnarray}
	&& \hspace{-15pt} \hat{H} = \! \sum\limits_{\bm k} \xi_{\bm k} \hat{N}_{\bm k} - V_b \hat{J}_z - \frac{\Delta_0}{2} \! \sum\limits_{\sigma = \pm} \hat{J}_\sigma \hat{\Phi}_\sigma + \omega_0 \hat{a}^\dagger \hat{a} + \frac{\omega_0}{2}, \label{H} \\
	&& \hspace{-15pt} \hat{N}_{\bm k} = \sum\limits_{s = \uparrow, \downarrow} \Psi_{\bm k, s}^\dagger \Psi_{\bm k, s} = \hat{n}_{\bm k} + N_{\bm k}, \label{Nk} \\
	&& \hspace{-15pt} \hat{J}_{\nu} = \sum\limits_{\bm k} \sum\limits_{s = \uparrow, \downarrow} \Psi_{\bm k, s}^\dagger \frac{\eta_\nu}{2} \Psi_{\bm k, s} = \hat{j}_\nu + J_\nu , \label{Jnu} \\
	&& \hspace{-15pt} \hat{\Phi}_\sigma = e^{\sigma g (\hat{a} - \hat{a}^\dagger)} = \hat{\phi}_\sigma + \Phi_\sigma , \label{Phi}
\end{eqnarray}
where $\bm k = (k_x, k_y)$ is the in-plane momentum of electrons, $\Psi_{\bm k, s}^T = (c_{\bm k, u, s}, c_{\bm k, d, s})$, $c_{\bm k, u, s}$ ($c_{\bm k, d, s}$) the electron field operator with momentum $\bm k$ and spin $s \in \{\uparrow, \downarrow\}$ corresponding to the upper (lower) minimum of the DQW, $\hat{N}_{\bm k}$ the electron number operator corresponding to the momentum $\bm k$, $\hat{J}_\nu$, $\nu \in \{z, \pm\}$, the dipole moment operator, $\eta_\nu$ the Pauli matrices acting on the $(u, d)$ subspace, $\eta_\pm = \eta_x \pm i \eta_y$, $\hat{\Phi}_\sigma$ corresponds to the Peierls substitution, $\hat{a}$ ($\hat{a}^\dagger$) the photon annihilation (creation) operator; $N_{\bm k}$, $J_\nu$, $\Phi_\sigma$ the ground state averages of corresponding operators; $\hat{n}_{\bm k}$, $\hat{j}_\nu$, $\hat{\phi}_\sigma$ the normal-ordered operators; $V_b$ the electrostatic bias between the upper and the lower quantum wells, $\Delta_0$ the hybridization of the DQW due to a finite overlap between the quantum wells, $\omega_0$ the bare cavity frequency, and $\xi_{\bm k}$ the electron dispersion, see Fig.~\ref{fig:1},
\begin{eqnarray}
	&& \xi_{\bm k} = \frac{k^2}{2 m} - E_F , \label{xik}
\end{eqnarray}
where $m$ is the effective mass, $E_F$ the Fermi energy.
The dimensionless coupling constant $g$ is defined as follows,
\begin{eqnarray}
	&& g = e A_z d = \sqrt{\frac{W}{\omega_0}} , \hspace{5pt} W = \frac{2 \pi e^2 d^2}{\varepsilon V_\text{eff}} . \label{g}
\end{eqnarray}
Here, $A_z$ is the vector-potential of the cavity mode polarized along the $z$ axis, see Fig.~\ref{fig:1}, $d$ the DQW width, $e$ the elementary charge, $\varepsilon$ the dielectric constant, and $V_\text{eff}$ the effective mode volume.
The cavity mode is almost completely localized in the capacitor, so $V_\text{eff}$ can be estimated as a capacitor volume, $V_\text{eff} = \Omega L_c$, where $\Omega$ is the sample area, $L_c$ the distance between the capacitor plates, see Fig.~\ref{fig:1}. {The mode is assumed to be uniform in the $xy$-plane, and thus it carries only a negligible small in-plane momentum due to finite size-effects.}
In this work, we consider the regime when two lowest electron subbands are partially occupied, meanwhile other subbands are split by the energy gap that is much larger than $E_F$.
Throughout the paper, we use CGS units and also set the Planck and Boltzmann constants to unity, $\hbar = k_\text{B} = 1$.

We are mostly interested in studying the {ultra-strong (and beyond)} coupling limit, $g \gtrsim 1$, that puts substantial constraints on the parameters of the system.
The capacitor volume is limited from below by the volume of DQW 2DEG, i.e., $V_\text{eff} > d \, \Omega$.
The area $\Omega$ of the 2DEG is also constrained from below as $\Omega \gtrsim d^2$, which puts the following upper bound on the coupling constant $g$,
\begin{eqnarray}
	&& g < \sqrt{\frac{\lambda_0}{d} \frac{e^2}{\varepsilon c}} \approx \sqrt{\frac{\lambda_0}{137 \varepsilon d}} , \label{gbound}
\end{eqnarray}
where $\lambda_0 = 2 \pi c /\omega_0$ is the wavelength of the cavity mode, and $e^2/c \approx 1/137$ the fine structure constant.
In semiconductor materials $\varepsilon \sim 10$.
Therefore, the ultra-strong coupling regime can be achieved if $\lambda_0 \gtrsim 10^3 d$ which corresponds to $g \gtrsim 1$.
We point out that such condition is impossible in the Fabry-P\'{e}rot resonator, where $\lambda_0 \sim 2 d$ is constrained by the capacitor dimensions.
However, $\lambda_0$ in the $LC$-cavity can be many orders of magnitude larger than the capacitor dimensions, allowing for ultra-strong light-matter coupling~\cite{ModeCompressionPRB, ModeCompressionNanoLett}. 

Separation of operators into the normal-ordered part and the ground-state average, see Eqs.~(\ref{Nk})--(\ref{Phi}), results in the following decomposition of the Hamiltonian $\hat{H}$,
\begin{eqnarray}
	&& \hspace{-10pt} \hat{H} = E_0 + \hat{H}_{\mathrm{e}} + \hat{H}_{\mathrm{ph}} + \hat{H}_{\mathrm{int}} , \label{Hsep} \\
	&& \hspace{-10pt} E_0 = \sum_{\bm k} \xi_{\bm k} N_{\bm k} - V_b  J_z , \label{E0} \\
	&& \hspace{-10pt} \hat{H}_{\mathrm{e}} =  \sum\limits_{\bm k} \xi_k \hat{n}_{\bm k} - V_b \hat{j}_z - \Delta \hat{j}_x , \label{e} \\
	&& \hspace{-10pt} \hat{H}_{\mathrm{ph}} = \omega_0 g^2 \hat{P}^2 + \hat{V}(\hat{Q}), \label{ph} \\
	&& \hspace{-10pt} \hat{V}(\hat{Q}) = \frac{\omega_0 \hat{Q}^2}{4 g^2} - \Delta_0 J_\parallel \cos \hat{Q} , \label{V} \\
	&& \hspace{-10pt} \hat{H}_{\mathrm{int}} = - \frac{\Delta_0}{2} \sum\limits_{\sigma = \pm} \hat{\phi}_\sigma \hat{j}_\sigma , \label{Hint}
\end{eqnarray}
where $\hat{H}_{\mathrm{e}}$ and $\hat{H}_{\mathrm{ph}}$ are the electron and the photon Hamiltonians, respectively, $\hat{H}_{\mathrm{int}}$ the interaction Hamiltonian, $E_0$ is the electron contribution to the ground-state energy, $\hat{Q}$ and $\hat{P}$ the canonical normal-ordered photon coordinate and momentum operators, $[\hat{P}, \hat{Q}] = -i$,
\begin{eqnarray}
	&& \hat{Q} = i g (a - a^\dagger), \hspace{5pt} \hat{P} = \frac{a + a^\dagger}{2 g} . \label{QP}
\end{eqnarray}
Parameters $J_\parallel$ and $\Delta$ are defined as follows,
\begin{eqnarray}
	&& J_\parallel = J_\pm , \hspace{5pt} \Phi_0 = \Phi_\pm = \langle e^{\mp i \hat{Q}} \rangle , \hspace{5pt} \Delta = \Delta_0 \Phi_0 , \label{Delta} 
\end{eqnarray}
where $J_\pm$ and $\Phi_\pm$ are defined in Eqs.~(\ref{Jnu}) and  (\ref{Phi}), $\Delta$ stands for the renormalized hybridization.

\section{Electron Hamiltonian}

The electron Hamiltonian in Eq.~(\ref{e}) is diagonalized by the following unitary transformation,
\begin{equation}
	U = \left(
	\begin{array}{cc}
		\cos \frac{\displaystyle \alpha}{\displaystyle 2} & -\sin \frac{\displaystyle \alpha}{\displaystyle 2} \\ [4pt]
		\sin \frac{\displaystyle \alpha}{\displaystyle 2}  & \cos \frac{\displaystyle \alpha}{\displaystyle 2}
	\end{array}
	\right) , \label{U}
\end{equation}
where first (second) column corresponds to the eigenstate $|+\rangle$ ($|-\rangle$) with the energy $\xi_{\bm k, +}$ ($\xi_{\bm k, -}$),
\begin{eqnarray}
	&& e^{i \alpha} = \frac{V_b + i \Delta}{E_g} , \label{thetatilde} \\
	&& E_g = \sqrt{V_b^2 + \Delta^2}, \label{Eg} \\
	&& \xi_{\bm k, \sigma} = \xi_{\bm k} - \sigma \frac{E_g}{2} , \label{tildexik}
\end{eqnarray}
where $\sigma = \pm 1$, $\xi_{\bm k}$ is defined in Eq.~(\ref{xik}), and $E_g$ is the renormalized splitting between first two occupied subbands.
The ground-state quantities $N_{\bm k}$, $J_z$, and $J_\parallel$ are found as
\begin{eqnarray}
	&& N_{\bm k} = \sum\limits_{\sigma = \pm 1} \frac{2}{e^{\beta \xi_{\bm k, \sigma}} + 1} , \label{Nkaverage}\\
	&& J_z = \frac{m \Omega V_b}{2\pi} , \hspace{5pt} J_\parallel = \frac{m \Omega \Delta}{2\pi} , \label{Spincondensates}
\end{eqnarray}
where $\Omega$ is the 2DEG area, the spin degeneracy is taken into account, $\beta = 1/T$, and we assume small temperatures $T \ll E_F - E_g/2$.

As our further discussion will be focused on heat capacity, we write down here the standard formula of the heat capacity of a 2DEG, and we note that it is linear in $T$ at $T \ll E_F$, see Ref.~\cite{AshcroftMermin}:
\begin{align}
	C_\text{2DEG} \approx N_F \Omega T\frac{\pi^2}{3} =  \frac{2\pi}{3} m \Omega T ,
\end{align}
where $N_F = 2 m/\pi$ is the density of states at the Fermi level.
Here, we take into account both subbands and the spin degeneracy of each subband. {We work in the regime where $T \ll E_F - E_g/2$, such that omitted terms are exponentially suppressed by the factor $e^{-\beta(E_F-E_g/2)}$.}

\section{Photon Hamiltonian}

The photon Hamiltonian contains the cosine term originating from the light-matter coupling, see Eqs.~(\ref{ph}) and  (\ref{V}).
The amplitude of the cosine term $\propto \Delta_0 J_\parallel \gg \Delta_0$ is proportional to $J_\parallel \gg 1$ that measures total number of hybridized electrons in the DQW.  
At large bare cavity frequency $\omega_0 \gg g^2 \Delta_0 J_\parallel$, the cosine term can be treated as a weak perturbation at any light-matter coupling $g$ because matrix elements of $\cos \hat{Q}$ do not exceed one in absolute value,
\begin{eqnarray}
	&& \hat{H}_{\mathrm{ph}} \approx \omega_0 \left(g^2 \hat{P}^2 + \frac{\hat{Q}^2}{4 g^2} \right) , \hspace{5pt} \omega_0 \gg g^2 \Delta_0 J_\parallel . \label{phweak}
\end{eqnarray}
In this limit, renormalizations of the LC cavity frequency as well as the coupling constant are negligible, and can be taken into account via usual perturbation theory.

\begin{figure}[!t]
  \includegraphics[width=0.98\columnwidth]{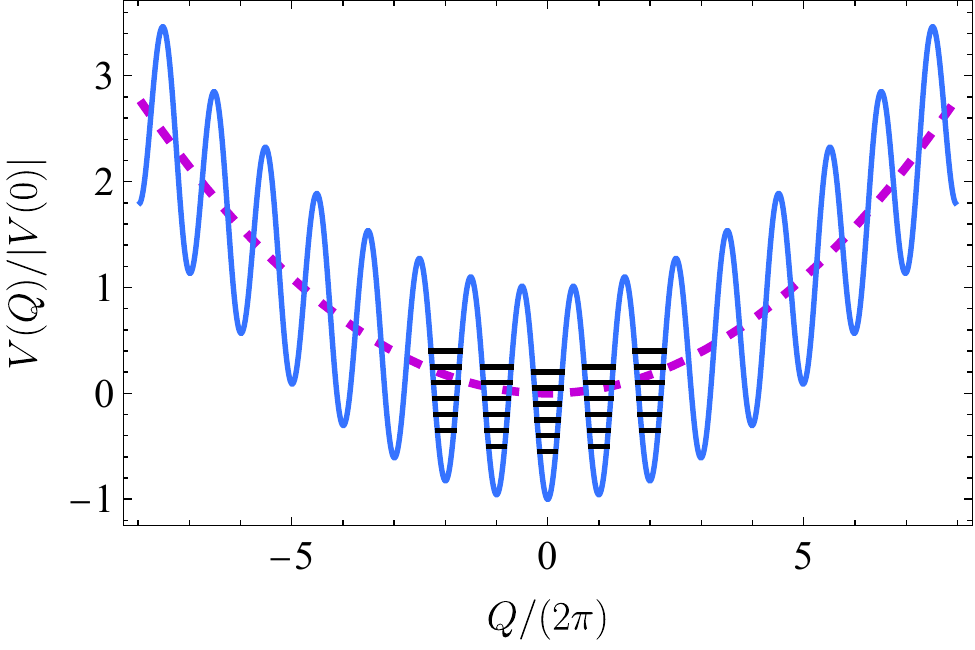}
  \caption{The schematic plot of the effective photonic potential $V(Q)$ [Eq.~(\ref{V})] at $\zeta \gg \zeta_c$ with large number of minima located at $Q \approx 2 \pi n$, $n$ is an integer.  {Here, $V(Q)$ consists of two terms: the bare cavity potential $\propto \omega_0 Q^2/g^2$ and the  term  $\propto\Delta_0 J_\parallel \cos{Q}$, originating from the cavity coupling to intersubband transitions. In mesoscopic systems $J_\parallel\gtrsim 10^3$, $\Delta_0\sim1$~meV, such that the amplitude of the cosine-term, $\Delta_0 J_\parallel$ is substantially greater than any other energy scale in the model. For this reason, the only optical transitions available are within a single cosine-minimum, each of which is characterized by large renormalized cavity frequency $\tilde{\omega}$ [see Eqs.~(\ref{omegan}) and (\ref{omega0condition})] and small renormalized coupling $\tilde{g}$ [see Eqs.~(\ref{gtilde}) and (\ref{gconstraint})]; optical transitions between different cosine-minima are almost impossible due to the very large amplitude $\Delta_0 J_\parallel$ of the cosine-term. On the other hand, the presence of multiple cosine-minima with offsets defined by the bare cavity potential is manifested in thermodynamic quantities. The corresponding energy scale is given by  $\omega_0/g^2$ of the bare cavity potential $\propto \omega_0 Q^2/g^2$. This is a characteristic energy scale of the emergent degree of freedom.} The harmonic oscillator levels are shown for a few local minima as a guide to the eye. 
  \label{fig_V_Q}}
  \end{figure}

In this paper, we concentrate on the opposite limit of small cavity frequency, $\omega_0 \ll g^2 \Delta_0 J_\parallel$.
In this limit the cosine term qualitatively changes the spectrum of photons due to emergent minima in the photonic potential $\hat{V}(\hat{Q})$, see Fig.~\ref{fig_V_Q}.
Semiclassical minima $Q_n$ of $\hat{V}(\hat{Q})$, $n$ is an integer, satisfy the following equation,
\begin{eqnarray}
	&& \sin Q_n = - \frac{Q_n}{\zeta}, \hspace{5pt} \cos Q_n > -\frac{1}{\zeta} , \label{Qminima} \\
	&& \zeta = \frac{2 g^2 \Delta_0 J_\parallel}{\omega_0} , \label{zeta}
\end{eqnarray}
where additional condition in Eq.~(\ref{Qminima}) ensures that $Q_n$ is a local minimum of $\hat{V}(\hat{Q})$.
If $\zeta < \zeta_c \approx 4.6$, $\hat{V}(\hat{Q})$ contains a single minimum at $Q = 0$. 
In this case, the photonic spectrum is slightly modified by the anharmonicity but it does not contain qualitatively new features.
However, at $\zeta > \zeta_c$ new minima emerge, see Fig.~\ref{fig_V_Q}.
In what follows, we analyze the regime when $\zeta \gg \zeta_c$ corresponding to large number of minima. 

{Here we focus on mesoscopic systems, such that the subband splitting $\Delta_0\sim1$~meV, and $J_\parallel\gtrsim 10^3$ (showing the imbalance of electrons between the subbands), hence the amplitude of the cosine-term $\Delta_0 J_\parallel $ dominates all the other energy scales in the problem. Therefore, the parameter $\zeta$ is naturally much greater than 1 for mesoscopic systems. From now on, we consider finite temperatures $T\ll \Delta_0 J_\parallel$.}

In strongly anharmonic regime $\zeta \gg \zeta_c \approx 4.6$, we can expand solutions of Eq.~(\ref{Qminima}) with respect to $1/\zeta \ll 1$,
\begin{eqnarray}
	&& \hspace{-10pt} Q_n \approx 2 \pi n , \hspace{5pt} -N_{\mathrm{max}} \le n \le N_{\mathrm{max}} , \label{Qn}\\
	&& \hspace{-10pt} N_{\mathrm{max}} \approx \left \lfloor{\frac{\zeta}{2 \pi} + \frac{1}{4}}\right \rfloor , \label{Nmax}
\end{eqnarray}
where $2 N_{\mathrm{max}} + 1$ is the total number of minima of $\hat{V}(\hat{Q})$, $\lfloor{x}\rfloor$ is the floor function. 
The global minimum always corresponds to $n = 0$ with $Q_0 = 0$.
In particular, the ground-state properties at zero temperature are unaffected by additional minima at $n \ne 0$.
However, these new minima contribute to thermodynamic properties of such anharmonic cavity at finite temperatures.

In the limit $\zeta \gg \zeta_c$ each minimum of $\hat{V}(\hat{Q})$ can be approximated by an oscillator (see Fig.~\ref{fig_V_Q}),
\begin{eqnarray}
	&& \hat{V}(Q_n + \hat{q}) \approx V(Q_n) + \frac{V''(Q_n)}{2} \hat{q}^2 , \label{Vmin}
\end{eqnarray}
where $\hat{q}$ is the photonic coordinate operator near the $n^{\mathrm{th}}$ minimum satisfying the Heisenberg commutation relation $[\hat{P}, \hat{q}] = -i$, $\hat{P}$ is defined in Eq.~(\ref{QP}).
Here, we take into account that $Q_n$ is a minimum of $V(Q)$, i.e. $V'(Q_n) = 0$ and $V''(Q_n) > 0$.
The renormalized frequency $\omega_n$ is approximately the same for each minimum (if $n$ is not too close to $\pm N_{\mathrm{max}}$),
\begin{eqnarray}
	&& \omega_n = \sqrt{2 \omega_0 g^2 V''(Q_n)} \approx \omega_0 \sqrt{1 + \zeta} \equiv \tilde{\omega} . \label{omegan}
\end{eqnarray}
Large barriers $\sim \Delta_0 J_\parallel$ between the minima allow us to neglect the tunneling and consider the contributions of each minimum to the photonic statistical sum $Z_{\mathrm{ph}}$ independently,
\begin{eqnarray}
	&& Z_{\mathrm{ph}} \approx e^{\beta \Delta_0 J_\parallel} Z_{\mathrm{osc}} Z_{\mathrm{b}} , \label{Zphapprox} \\
	&& Z_{\mathrm{osc}} = \frac{1}{2} {\mathrm{csch}}\left(\beta \frac{\tilde{\omega}}{2}\right) , \label{Zosc} \\
	&& Z_{\mathrm{b}} = \sum\limits_{n = -N_{\mathrm{max}}}^{N_{\mathrm{max}}} e^{-\beta \mathcal{E}_n} \approx \theta_3(\mu) , \label{Zboson} \\
	&& \mathcal{E}_n = V(Q_n) + \Delta_0 J_\parallel \approx \omega_0 \frac{\pi^2 n^2}{g^2}, \label{En} \\
	&& \mu = \exp\left(-\frac{\beta \mathcal{E}_n}{n^2}\right) \approx \exp\left(-\frac{\pi^2}{g^2} \beta \omega_0\right) , \label{mu}
\end{eqnarray}
where $\mathcal{E}_n$ is the energy offset of the $n^{\mathrm{th}}$ minimum, $Z_{\mathrm{osc}}$ is the statistical sum of a harmonic oscillator, $Z_{\mathrm{b}}$ is the statistical sum of an emergent degree of freedom, $\theta_3(\mu) = \sum_{\mathbb{Z}} \mu^{n^2}$ is the Jacobi Theta  function, $\mathbb{Z}$ stands for integers.
We approximately set $N_{\mathrm{max}} \to \infty$ in Eq.~(\ref{Zboson}), which is valid if $T \ll \mathcal{E}_{N_{\mathrm{max}}} \approx W (\Delta_0 J_\parallel/\omega_0)^2$, where we used Eqs.~(\ref{g}) and (\ref{Nmax}).
We are interested in $T \sim \mathcal{E}_1 \ll \mathcal{E}_{N_{\mathrm{max}}}$, which justifies this approximation.

Non-equidistant $\mathcal{E}_n \propto n^2$ spectrum of the minima offsets can be effectively described by a particle, trapped in the infinite square-well potential.
This emergent degree of freedom does not couple to the current operator $e d \, \delta \hat{H}/\delta \hat{Q}$ because $\hat{V}'(Q_n) = 0$ and $e^{\pm i Q_n} = 1 + \mathcal{O}(1/\zeta) \approx 1$. 
Therefore, the current operator takes the same form near each minimum, hence the current-current correlator takes the same value as if there is only one harmonic minimum with the renormalized frequency $\tilde{\omega}$.  Weak $1/\zeta$ coupling is possible due to $\hat{H}_{\mathrm{int}}$ and that $e^{\pm i Q_n} \approx 1 \mp i Q_n/\zeta$, where we take into account $1/\zeta$ correction to $Q_n$, see Eq.~(\ref{Qminima}).
However, this correction contributes only to $1/\zeta^2$ term in the current-current correlator which is negligible.
The absorption spectrum is proportional to the imaginary part of the $\langle \hat{Q} \hat{Q} \rangle$ correlator.
Near the n$^{\mathrm{th}}$ minimum, $\hat{Q} = Q_n + \hat{q}$, so $\langle \hat{Q} \hat{Q} \rangle = Q_n^2 + \langle \hat{q} \hat{q} \rangle$.
As $Q_n^2$ does not contribute to the imaginary part, then the emergent degree of freedom is not visible in the absorption spectrum.
Therefore, the emergent degree of freedom can only be detected via the heat capacity measurement.

Let us analyze physical conditions leading to Eq.~(\ref{Zphapprox}).
This approximation is valid if the minima of $\hat{V}(\hat{Q})$ contain large number of harmonic oscillator levels, i.e. $\Delta_0 J_\parallel \gg \tilde{\omega}$. 
Together with the condition $\zeta \gg 1$ providing large number of minima, see Eq.~(\ref{Nmax}), this regime corresponds to the following constraint,
\begin{eqnarray}
	&& \frac{\omega_0}{\Delta_0 J_\parallel} \ll \mathrm{min}\left(g^2, \frac{1}{g^2}\right) . \label{condition}
\end{eqnarray}
We point out that $J_\parallel$ depends on the renormalized hybridization $\Delta$, see Eqs.~(\ref{Delta}) and (\ref{Spincondensates}).
Neglecting effects of $H_{\mathrm{int}}$, we find the hybridization renormalization factor $\Phi_0$, see Eq.~(\ref{Delta}), which is analogue of the Debye-Waller factor, 
\begin{eqnarray}
	&& \Phi_0 \approx \exp\left[-\tilde{g}^2 \left(N_{\mathrm{ph}} + \frac{1}{2}\right)\right] , \label{Phi0} \\
	&& \tilde{g} = g  \sqrt{\frac{\omega_0}{\tilde{\omega}}} = \frac{g}{(\zeta + 1)^{\frac{1}{4}}} , \label{gtilde} \\
	&& N_{\mathrm{ph}} = \frac{1}{e^{\beta \tilde{\omega}} - 1} = \frac{1}{2} \left[\coth\left(\frac{\beta \tilde{\omega}}{2}\right) - 1\right], \label{Nph}
\end{eqnarray}
where $\tilde{g}$ stands for the renormalized light-matter coupling, $N_{\mathrm{ph}}$ is average number of thermal photons in the harmonic oscillator with frequency $\tilde{\omega}$.
Here, we used Eqs.~(\ref{Qn}) and  (\ref{omegan}) and the quadratic expansion of $\hat{V}(\hat{Q})$ near each minimum, see Eq.~(\ref{Vmin}).
As $e^{\pm i Q_n} \approx 1$, see Eq.~(\ref{Qn}), the operator $e^{\pm i (Q_n + \hat{q})} \approx e^{\pm i \hat{q}}$ takes the same form near each minimum.
Therefore, $\Phi_0$ is approximately equal to the harmonic oscillator value.
We point out that in this regime $\tilde{g}^2 \tilde{\omega} = g^2 \omega_0 = W$, i.e. the energy scale $W$ remains unrenormalized, see Eq.~(\ref{g}).
The coupling to 2DEG results in heavier oscillator frequency $\tilde{\omega} \gg \omega_0$, see Eq.~(\ref{omegan}), which, in turn, strongly reduces the effective light-matter coupling, $\tilde{g} \ll g$, see Eq.~(\ref{gtilde}).
The renormalized coupling is small, $\tilde{g} \ll 1$, as soon as $W \ll \Delta_0 J_\parallel$, which corresponds to the following condition on the hybridization,
\begin{eqnarray}
	&& \Delta_0 \gg \frac{2 \pi}{m \Omega} \sqrt{\frac{d^2}{L_c a_B}} , \label{Deltacond} \\
	&& a_B = \frac{\varepsilon}{m e^2} , \label{Bohr}
\end{eqnarray} 
where $a_B$ is the effective Bohr radius.
The energy scale $2 \pi/(m \Omega)$ corresponds to the discretization of the in-plane energy levels due to the finite size $\Omega$ of the 2DEG.
As the distance between the capacitor plates is greater than the DQW width, $L_c > d$, and $d/a_B < 100$ in realistic devices ($a_B \gtrsim 1\,$nm, $d \lesssim 100\,$nm), the condition Eq.~(\ref{Deltacond}) is always satisfied in a mesoscopic tunnel-coupled DQW 2DEG.
Therefore, $\Phi_0 \approx 1$, i.e. the renormalization of the DQW hybridization is not essential, $\Delta \approx \Delta_0$.
Using Eqs.~(\ref{g}), (\ref{Spincondensates}), and (\ref{Deltacond}), the condition Eq.~(\ref{condition}) corresponds to the following constraint on the cavity frequency,
\begin{eqnarray}
	&& \omega_0 \ll \tilde{\omega} \approx \omega_0 \sqrt{\zeta} \approx \Delta_0 \sqrt{\frac{2 d^2}{L_c a_B}} . \label{omega0condition}
\end{eqnarray}
The renormalized oscillator frequency $\tilde{\omega}$ in this regime is no longer dependent on the bare cavity frequency $\omega_0$ as it rather represents a {polariton}.
The condition given by Eq.~(\ref{omega0condition}) is equivalent to the following constraint for the light-matter coupling constant $g$,
\begin{eqnarray}
	&& g \gg \tilde{g} \approx \sqrt{\frac{2 \pi}{m \Omega \Delta_0} \sqrt{\frac{d^2}{2 L_c a_B}}} . \label{gconstraint}
\end{eqnarray}
As $\tilde{g} \ll 1$, which follows from Eq.~(\ref{Deltacond}), the light-matter coupling $g$ can still be much smaller than one, i.e. the {ultra-strong (and beyond)} coupling $g \gtrsim 1$ is not even required for the considered regime.
Therefore, Eqs.~(\ref{Deltacond}) and  (\ref{omega0condition}) provide necessary and sufficient conditions at which the effective photonic potential $\hat{V}(\hat{Q})$ contains large number of harmonic oscillator minima.

\begin{figure}[h]
	\includegraphics[width=0.98\columnwidth]{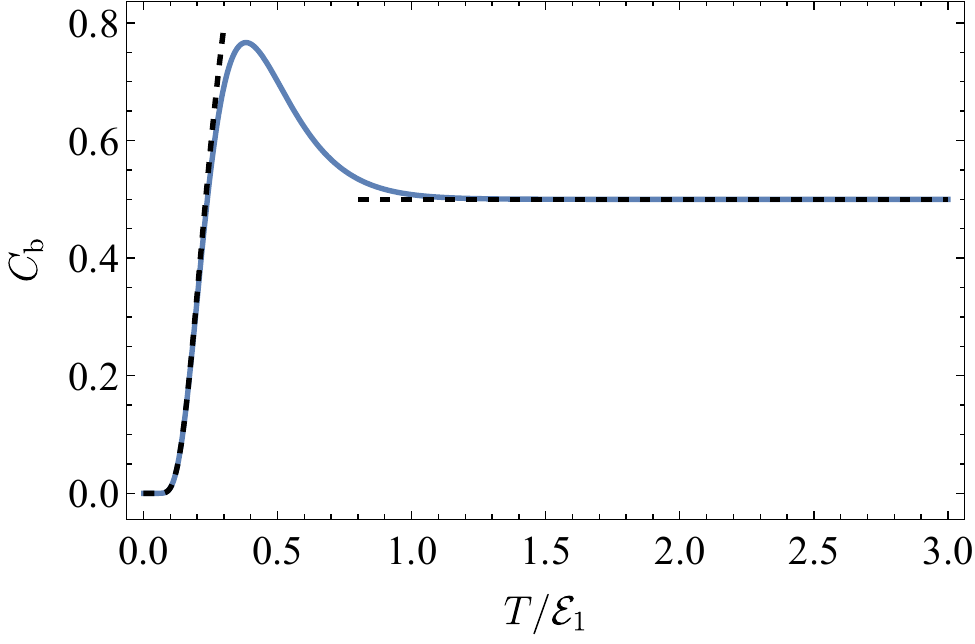}
	\caption{
		The contribution to the heat capacity from the emergent degree of freedom $C_{\mathrm{b}}(T/\mathcal{E}_1)$ as a function of temperature: the exact result (solid curve) given by Eq.~(\ref{eq:C_ph_and_C_b}) compared to the asymptotic expressions (dashed curves) given in Eq.~(\ref{eq:C_b_approx}).
		The Schottky anomaly is located at $T^* \approx 0.38 \, \mathcal{E}_1$ with the value $C_b^* \approx 0.77$.
		\label{fig_C_b} }
	\vspace{-5pt}
\end{figure}

Next, we analyze the thermodynamic properties of this system in the regime when Eqs.~(\ref{Deltacond}) and  (\ref{omega0condition}) are satisfied.
Photonic free energy $\mathcal{F}_{\mathrm{ph}} = - T \ln Z_{\mathrm{ph}}$ then follows from Eq.~(\ref{Zphapprox}),
\begin{eqnarray}
	&& \mathcal{F}_{\mathrm{ph}} \approx -\Delta_0 J_\parallel + \mathcal{F}_{\mathrm{osc}} + \mathcal{F}_{\mathrm{b}} , \label{Fph} \\
	&& \Delta_0 J_\parallel \approx \frac{m \Omega \Delta_0^2}{2 \pi} - \frac{\tilde{\omega}}{2} \left(N_{\mathrm{ph}} + \frac{1}{2}\right) , \label{Peierlsenergy} \\
	&& \mathcal{F}_{\mathrm{osc}} = T \ln \left[2 \sinh\left(\beta \frac{\tilde{\omega}}{2}\right)\right] , \label{Fosc} \\
	&& \mathcal{F}_{\mathrm{b}} = - T \ln \left[\theta_3(\mu) \right].
\end{eqnarray}
In Eq.~(\ref{Peierlsenergy}), we used Eqs.~(\ref{Delta}), (\ref{Spincondensates}), and (\ref{Phi0}), where $\Phi_0$ is expanded up to $\tilde{g}^2$ order as $\tilde{g} \ll 1$.
We also used approximate Eq.~(\ref{omega0condition}) for $\tilde{\omega}$.
Terms that are neglected in Eq.~(\ref{Peierlsenergy}) are proportional to $\tilde{g}^2 \tilde{\omega} = W \ll \tilde{\omega}$ which is of the order of a finite-size discretization of the 2DEG, $2 \pi/(m \Omega)$, see Eq.~(\ref{g}) for definition of $W$.
The heat capacity corresponding to the photonic Hamiltonian $\hat{H}_{\mathrm{ph}}$ is then the following, 
\begin{eqnarray}
	&& C_{\mathrm{ph}} = \frac{y^2}{\sinh^2 (y)} \left[2 - y \coth (y)\right] + C_{\mathrm{b}}, \hspace{5pt} y = \frac{\beta \tilde{\omega}}{2} , \label{Cph} \\
	&& C_{\mathrm{b}} = \ln^2 (\mu) \left(\mu \frac{\partial}{\partial \mu}\right)^2 \ln \theta_3 (\mu) \, ,\label{eq:C_ph_and_C_b} 
\end{eqnarray}
where $\mu$ is given by Eq.~(\ref{mu}).
Here, $C_{\mathrm{b}}$ is the contribution to the heat capacity from the emergent degree of freedom.
Asymptotic values of $C_{\mathrm{b}}$ are the following,
\begin{equation}
	C_{\mathrm{b}} \to \left\{
	\begin{array}{cc}
		2 e^{-\beta \mathcal{E}_1} \left(\beta \mathcal{E}_1\right)^2 , & T \ll \mathcal{E}_1 \\[3pt]
		\frac{\displaystyle 1}{\displaystyle 2} , & T \gg \mathcal{E}_1 , 
			\label{eq:C_b_approx}
	\end{array}
	\right.
\end{equation}
where $\mathcal{E}_1 = \pi^2 \omega_0 /g^2$.  In Fig.~\ref{fig_C_b}, $C_{\mathrm{b}}$ is shown as a function of $T$ where it tends to zero at small temperatures $T \ll \mathcal{E}_1$ and tends to fixed value $C_{\mathrm{b}} = 1/2$ at $T \gg \mathcal{E}_1$.
At the crossover temperatures $T \sim \mathcal{E}_1$, $C_{\mathrm{b}}$ exhibits the Schottky anomaly, the peak at $T^* \approx 0.38 \, \mathcal{E}_1$ with the maximal value $C^*_{\mathrm{b}} \approx 0.77$.

\section{Interaction correction to the heat capacity}

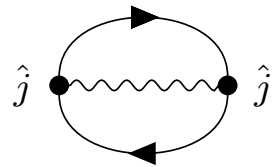
\begin{figure}[!b]
\[\vcenter{\hbox{\scalebox{1.6}{\begin{tikzpicture}
  \begin{feynman}
    \vertex[dot] (a) {};
     \node[left=0.3cm of a] (a0){$\hat{j}$};
    \vertex[right=1.4cm of a, dot] (b) {};
     \node[right=0.3cm of b] (b1){$\hat{j}$};
    \diagram*{
      (a)-- [photon] (b),
      (a) -- [fermion, with arrow=0.5, half left, looseness=1.2] (b),
      (b) -- [fermion, half left, looseness=1.2] (a),
    };
  \end{feynman}
\end{tikzpicture}}}}\]
 \centering
  \caption{The Feynman diagram showing the leading interaction contribution to the free energy, see Eq.~(\ref{Fint}). Black solid lines correspond to the electron Green's function, the wavy line stands for the propagator of $\hat{\phi}$ field, see Eq.~(\ref{Phi}), $\hat{j}$ stands for the dipole moment operator, see Eq.~(\ref{Jnu}).}
  \label{fig_feyn_fi}
\end{figure}

\begin{figure}[!tb]
	\includegraphics[width=0.98\columnwidth]{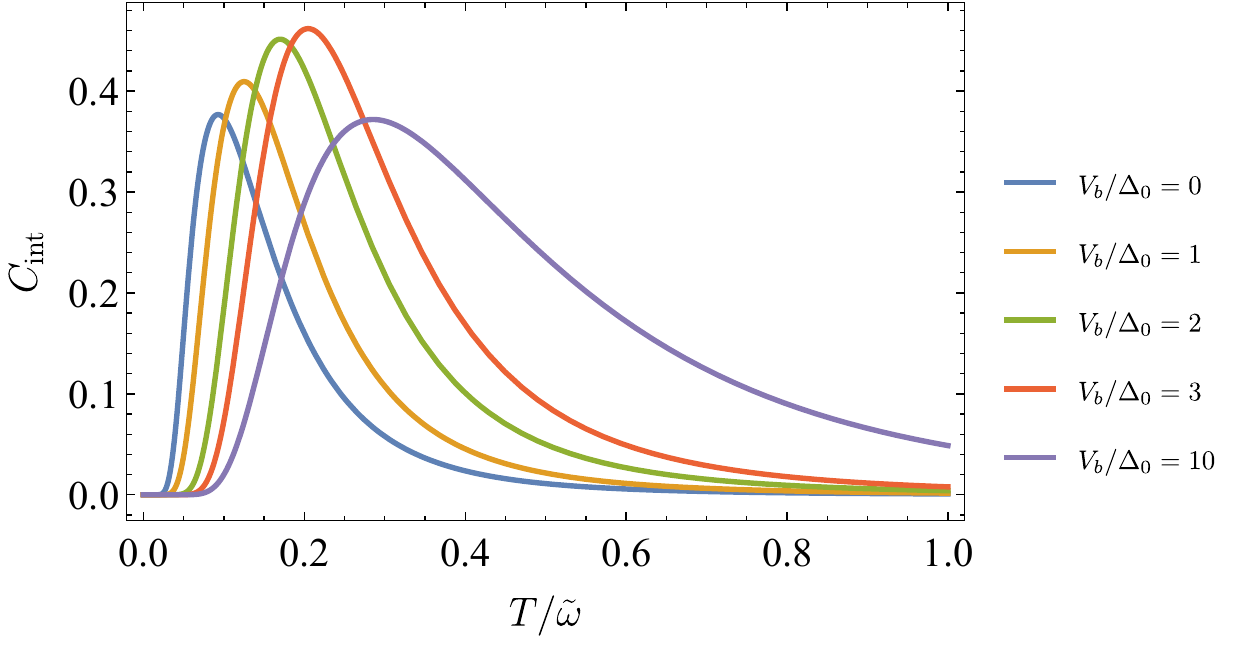}
	\caption{The $C_\text{int}$ contribution to the heat capacity vs. $T/\tilde{\omega}$ calculated at $\Delta_0 = 0.3 \, \tilde{\omega}$ and different values of $V_b/\Delta_0 = 0, 1, 2, 3, 10$. We choose the ratio $d^2/(a_B L_c) = 5$, which corresponds to parameters we later use in Sec.~\ref{sec:disc}.	\label{fig_C_int_diff_Vb}}
	\vspace{-5pt}
\end{figure}

Having found the Schottky anomaly at low temperature, we now proceed to analyzing the effect of the interaction term on it. In order to calculate the interaction contribution from $\hat{H}_{\mathrm{int}}$ to the heat capacity, we evaluate the interaction correction to the grand canonical potential,
\begin{align}	
& \mathcal{F}_{\mathrm{int}} = - T \ln \langle e^{-\int \hat{H}_{\mathrm{int}}d\tau}\rangle .
\end{align}
As the renormalized coupling constant is small, $\tilde{g} \ll 1$, we can expand this expression up to $\tilde{g}^2$ terms which is represented by the diagram in Fig.~\ref{fig_feyn_fi},
\begin{align}	
	& \mathcal{F}_{\mathrm{int}} \approx - \frac{T}{2} \int d\tau_1 d\tau_2\langle \mathcal{T}\{\hat{H}_{\mathrm{int}}(\tau_1)\hat{H}_{\mathrm{int}}(\tau_2)\} \, , \label{Fint}
\end{align}
where $\mathcal{T}$ stands for the time ordering operator.
This correction is proportional to $\tilde{g}^2 \Omega$, where 
$\tilde{g}^2 \propto 1/\Omega$ which follows from Eqs.~(\ref{g}) and (\ref{gtilde}), where $\zeta$ is independent of $\Omega$.
Therefore, the $\tilde{g}^2$ interaction correction given by Eq.~(\ref{Fint}) does not scale with the 2DEG area $\Omega$.
Higher-order contributions are proportional to at least $\Omega \tilde{g}^4 \propto 1/\Omega$ and we can safely neglect them.
Calculation of $\mathcal{F}_{\mathrm{int}}$ is outlined in Appendix~\ref{appendix_int}, here we provide the result,
\begin{eqnarray}
&& \hspace{-25pt} \mathcal{F}_{\mathrm{int}} \! \approx \! \frac{m \Omega}{4 \pi} \frac{\tilde{g}^2 \Delta_0^2 E_g}{E_g^2 - \tilde{\omega}^2} \!\! \left[\tilde{\omega} \coth\left(\!\frac{\beta E_g}{2}\!\right) \! - \! E_g \coth\left(\!\frac{\beta \tilde{\omega}}{2}\!\right)\! \right]. \label{Fintexplicit} 
\end{eqnarray}
The interaction correction to the heat capacity takes the following form,
\begin{eqnarray}
	&& C_{\mathrm{int}} = \frac{d^2}{2 a_B L_c} \frac{\Delta_0^2 E_g^2 \beta^2}{E_g^2 - \tilde{\omega}^2} \left[f(E_g) - f(\tilde{\omega})\right] , \label{Cint} \\
	&& f(E) \equiv \left[1 - \frac{\beta E}{\sinh(\beta E)}\right] \coth^2 \left(\frac{\beta E}{2}\right) .
\end{eqnarray}
Here, we take into account that $\tilde{g}^2 \tilde{\omega} = W$, where $W$ is defined in Eq.~(\ref{g}).
We point out that there is no singularity in $C_{\mathrm{int}}$ at $E_g = \tilde{\omega}$.
We plot $C_{\mathrm{int}}$ as a function of $T$ in Fig.~\ref{fig_C_int_diff_Vb} at fixed $\Delta_0 = 0.3 \, \tilde{\omega}$ and different values of bias $V_b/\Delta_0 = 0, 1, 2, 3, 10$. 
Generally, $C_{\mathrm{int}}$ demonstrates the asymmetric bell-shape as a function of $T$ with exponential decay at $T \ll \mathrm{min}(E_g, \tilde{\omega})$ and $C_{\mathrm{int}} \propto 1/T^4$ at $T \gg \mathrm{max}(E_g, \tilde{\omega})$.
Increasing the bias $V_b$ at $\Delta_0 < \tilde{\omega}$ shifts the low-temperature shoulder of $C_{\mathrm{int}}$ to higher temperatures: this behavior saturates at $E_g \gg \tilde{\omega}$, see Fig.~\ref{fig_C_int_diff_Vb}.

\section{Discussion and conclusions}
\label{sec:disc}

\begin{figure}[h]
	\includegraphics[width=0.98\columnwidth]{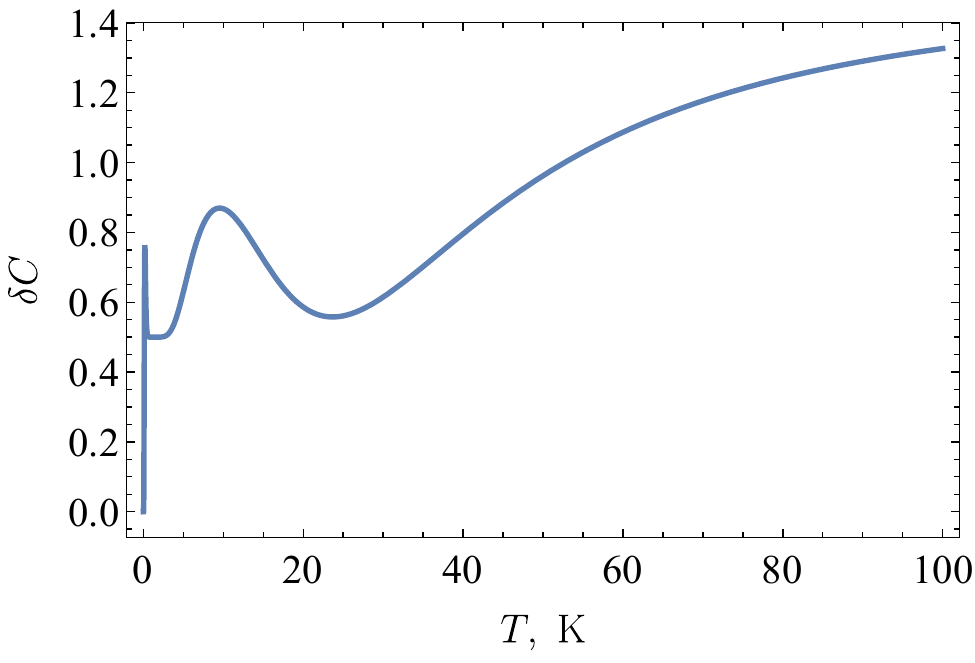}\put(-240,155){(a)}\\
	\includegraphics[width=0.98\columnwidth]{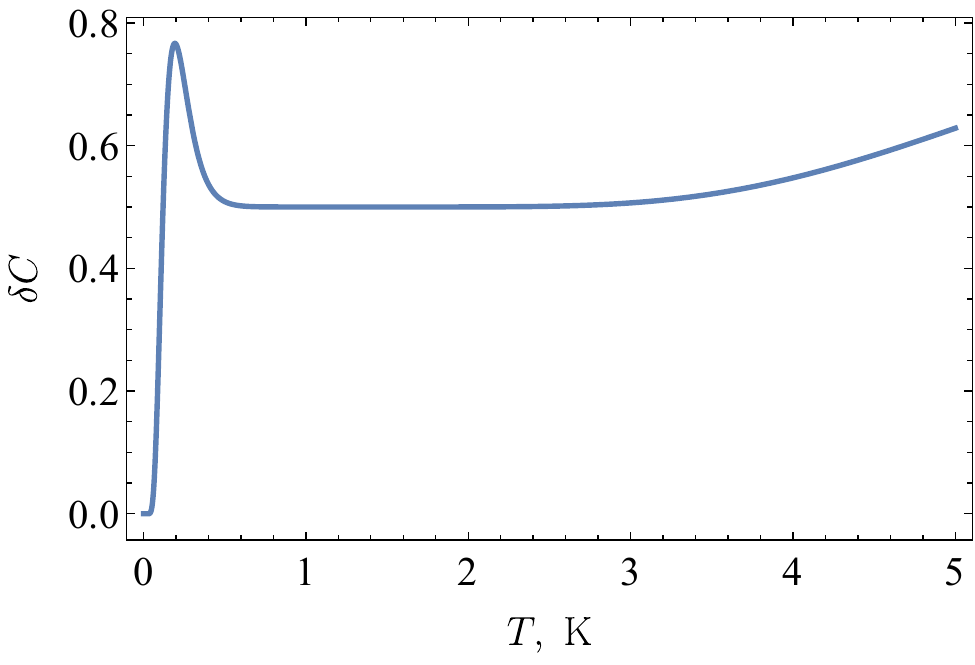}\put(-240,155){(b)}\\
	\caption{
		(a) The cavity-related contribution to the heat capacity $\delta C=C_\text{ph}+C_\text{int}$, see Eqs.~(\ref{Cph}) and (\ref{Cint}), calculated at the parameters chosen in Sec.~\ref{sec:disc}. The smooth curve at $T > 40\,$K corresponds to the oscillator contribution, it saturates at $T \sim \tilde{\omega} \sim 100\,$K.
		The wide peak at $T \approx 10\,$K corresponds to the interaction contribution $C_{\mathrm{int}}$.
		The main feature here is the Schottky-peak at low temperature $T \sim 0.2\,$K followed by the $0.5$ plateau (in units of $k_\text{B}$) heralding the emergence of the degree of freedom.
		The panel (b) shows $\delta C$ at very low temperatures highlighting the $0.5$ plateau and the Schottky anomaly. It is clearly seen that the Schottky peak is well separated from the interaction and oscillator contributions.
		\label{fig_heat_cap} }
	\vspace{-5pt}
\end{figure}

The half-step heat capacity contribution from the emergent degree of freedom has to be separated from large linear in $T$ heat capacity of the 2DEG and from $T^3$ heat capacity of the lattice which constitutes a difficult experimental challenge.
In order to increase accuracy of measurements, we suggest to use $N \gg 1$ arrays of 2DEGs coupled to individual $LC$ cavities.
Subtracting the linear-in-$T$ and $T^3$ contributions for each array and averaging the subtracted signal over $N$ arrays leads to substantial reduction of the noise by factor $1/\sqrt{N}$. 
Resulting measurement corresponds to $\delta C = C_{\mathrm{ph}} + C_{\mathrm{int}}$.
Bare cavity frequency $\omega_0$ can be measured when the 2DEG is completely depleted. 
{In addition to that,} a way of measuring the cavity-modified heat capacitance was recently proposed in Ref.~\cite{enkner2024enhancedfractionalquantumhall} where the split-ring resonators can be tuned close to or away from the 2DEG.
Thus, the heat capacity measurement can be performed on the same system with and without cavity.
Subtracting the heat capacity of the uncoupled system from the signal of the cavity-coupled system yields the measurement of $\delta C = C_{\mathrm{ph}} + C_{\mathrm{int}}$. Also, recent progress in the spectroscopy of a bilayer graphene ~\cite{adam2024entropyspectroscopybilayergraphene} shows experimental measurements of the entropy changes as small as fractions of $k_\text{B}$ exploiting the Maxwell relation.

We suggest to use $1\mu\text{m} \times 1\mu\text{m}$ semiconductor DQWs of the width $d = 50\,$nm, with effective mass $m = 0.1 m_e$ and the dielectric constant $\varepsilon = 10$.
The hybridization in a wide semiconductor DQW can still be sizable, we choose $\Delta_0 = 3\,$meV.
The bias $V_b$ is not very important here, so we choose $V_b = 0$.
The distance between the capacitor plates $L_c = 100\,$nm.
As we are interested in the heat capacity at $T \sim \mathcal{E}_1 \propto \omega_0^2/W$, it is essential to consider GHz cavities.
Here, we choose $\omega_0 = 0.01\,$meV.
At these parameters, we find $W \approx 23\,\mu$eV, $g \approx 1.5$, $J_\parallel \approx 650$, $\zeta \approx 0.9 \times 10^6$, $\tilde{\omega} \approx 9.5 \,$meV, $\tilde{g} \approx 0.05$, $\mathcal{E}_1 \approx 43\,\mu$eV. 
The temperature corresponding to the Schottky anomaly, see Fig.~\ref{fig_C_b}, $T^* \approx 0.2\,$K.
The 2DEG contribution at $T = T^*$ is not that overwhelming, $C_{\mathrm{2DEG}} \approx 9.2$.
The lattice heat capacity for the sample of considered dimensions can be estimated from the Debye model which gives~\cite{kittel2004} $C_{\mathrm{Deb}}\approx(12\pi^4/5)N_a k_\text{B} (T/\theta_D)^3 \approx 15$ for Si with the Debye temperature $\theta_D = 636\,$K.
From these estimates, we can conclude that such measurement can already be performed in realistic devices.

After successful subtraction of the 2DEG and the lattice contribution, the remaining heat capacity is given by the photon and interaction contributions, $\delta C = C_{\mathrm{ph}} + C_{\mathrm{int}}$.
We plot $\delta C$ as a function of $T$ in Fig.~\ref{fig_heat_cap} for chosen parameters.
Indeed, we find the emergent degree of freedom's contribution at low temperatures $T \sim T^* \approx 0.2\,$K.
The interaction contribution results in a bump at $T \sim 0.3 \Delta_0 \approx 10\,$K.
The oscillator step emerges at very high temperatures $T \sim \tilde{\omega} \sim 100\,$K. 
As we have shown, for the chosen parameters the Schottky peak remains visible against the background of the electron and phonon contributions as it manifests itself at $T < 1\,$K. However, the harmonic oscillator and interaction contributions which appear at $T \gg 1\,$K where the 2DEG and lattice contributions dominate, are difficult to resolve.
In contrast, the emergent degree of freedom's contribution corresponds to low temperatures $T \sim T^* \approx 0.2\,$K, where the 2DEG and the lattice contributions are still reasonably small.
We point out that the Schottky anomaly at $T = T^* = 0.38 \, \mathcal{E}_1$ provides the way of extracting the coupling constant $g$,
\begin{eqnarray}
	&& g = \pi \sqrt{0.38 \frac{\omega_0}{T^*}}, 
\end{eqnarray}
which in its turn is linked to the effective mode volume $V_\text{eff}$, see Eq.~(\ref{g}).

In conclusion, we have shown emergence of the effective degree of freedom in a cavity-coupled DQW 2DEG. The emergent degree of freedom can not couple to electromagnetic fields. Its presence can be measured via the heat capacity demonstrating the $0.5 k_B$ step at low temperatures $T \gg T^* \propto \omega_0/g^2$ and the Schottky anomaly at $T = T^*$.
Determination of the experimental value of $T^*$ provides the way to estimate the light-matter coupling constant $g$ and, correspondingly, the effective mode volume $V_\text{eff}$.

\begin{acknowledgments}
V.K.K. and D.M. acknowledge the support from the Georg H. Endress foundation. This project has received funding from the Swiss State Secretariat for Education, Research and Innovation (SERI) under the ERC replacement scheme following the discontinued participation of Switzerland to Horizon Europe. This work was also supported as a part of NCCR SPIN, a National Centre of Competence in Research, funded by the Swiss National Science Foundation (grant number 225153).
\end{acknowledgments}

\appendix 

\section{The interaction contribution to the heat capacity}
\label{appendix_int}

The interaction contribution to the grand canonical potential given by Eq.~(\ref{Fint}) can be represented as follows,
\begin{eqnarray}
	&& \mathcal{F}_{\mathrm{int}} = - \frac{\Delta_0^2}{8} \sum\limits_{\nu_1, \nu_2} \int\limits_0^{\beta} \Pi_{\nu_1, \nu_2}(\tau) D_{\nu_1, \nu_2} (\tau) \, d\tau , \label{Fapp} \\
	&& \Pi_{\nu_1, \nu_2}(\tau) = \left\langle \mathcal{T} \left\{ \hat{j}_{\nu_1} (\tau) \hat{j}_{\nu_2} (0) \right\} \right\rangle  , \label{Pi} \\
	&& D_{\nu_1, \nu_2} (\tau)  = \left\langle \mathcal{T} \left\{ \hat{\phi}_{\nu_1} (\tau) \hat{\phi}_{\nu_2} (0) \right\} \right\rangle , \label{Dapp}
\end{eqnarray}
where we just substituted Eq.~(\ref{Hint}) into Eq.~(\ref{Fint}) and used that the time-ordered average in Eq.~(\ref{Fint}) depends only on the time difference $\tau = \tau_1 - \tau_2$.
Indices $\nu_1$, $\nu_2$ take values $\pm 1$.

In order to calculate $D_{\nu_1, \nu_2} (\tau)$, we represent the photon coordinate as $\hat{Q} = Q_n + \hat{q}$ near the $n^{\mathrm{th}}$ minimum, where the photonic potential is harmonic, see Eq.~(\ref{Vmin}), with frequency $\tilde{\omega}$ that is independent of $n$, see Eq.~(\ref{omegan}). 
As $Q_n \approx 2 \pi n$, see Eq.~(\ref{Qn}), the operator $\hat{\phi}_\nu \approx :e^{-i \nu \hat{q}}: \equiv e^{-i \nu \hat{q}} - \Phi_0$ takes the same form near each minimum.
Notation $:\hat{O}:$ stands for normal ordering of an operator $\hat{O}$.
Therefore, $D_{\nu_1, \nu_2} (\tau)$ is the same as for a single oscillator with frequency $\tilde{\omega}$,
\begin{eqnarray}
	&& \hspace{-5pt} D_{\nu_1, \nu_2} (\tau) = \sum\limits_{n} \frac{e^{- \beta \mathcal{E}_n}}{Z_{\mathrm{b}}}  \left\langle \mathcal{T} \left\{ :e^{-i \nu_1 \hat{q}(\tau)}: :e^{-i \nu_2 \hat{q}(0)}: \right\} \right\rangle_n \nonumber \\
	&& \hspace{38pt} = \left\langle \mathcal{T} \left\{ :e^{-i \nu_1 \hat{q}(\tau)}: :e^{-i \nu_2 \hat{q}(0)}: \right\} \right\rangle_0 , \label{D1}
\end{eqnarray}
where $Z_{\mathrm{b}}$ is given by Eq.~(\ref{Zboson}), $\langle \dots \rangle_n$ stands for the statistical averaging near the $n^{\mathrm{th}}$ minimum of $\hat{V}(\hat{Q})$,  $\langle \dots \rangle_0$ stands for the averaging near the minimum at $Q = 0$.
Following the same line of reasoning, we also find that $\Phi_0$ is given by the average near the oscillator minimum at $Q = 0$, see Eq.~(\ref{Phi0}).
In order to evaluate the average in Eq.~(\ref{D1}), we first perform the canonical transformation of the photon Hamiltonian near $Q = 0$ minimum,
\begin{eqnarray}
	&& \hspace{-10pt} \hat{X} = \frac{\hat{q}}{\sqrt{2 \tilde{g}^2}} = \frac{\hat{b} + \hat{b}^\dagger}{\sqrt{2}} ,  \hspace{5pt} \hat{\mathcal{P}} = \sqrt{2 \tilde{g}^2} \hat{P} = \frac{\hat{b} - \hat{b}^\dagger}{i \sqrt{2}} , \label{Bop}
\end{eqnarray}
where $\tilde{g}$ is the renormalized coupling constant, see Eq.~(\ref{gtilde}), $\hat{b}$ ($\hat{b}^\dagger$) the annihilation (creation) operator of a photon in the $Q = 0$ minimum.
The photon Hamiltonian near $Q = 0$ minimum takes the form $\hat{H}_{\mathrm{ph}} (Q = 0) = \tilde{\omega} (b^\dagger b + 1/2)$.
From this, the interaction representation for $\hat{b}$ and $\hat{b}^\dagger$ is the following,
\begin{eqnarray}
	&& \hat{b}(\tau) = e^{- \tilde{\omega} \tau} \hat{b}, \hspace{5pt} \hat{b}^\dagger (\tau) = e^{\tilde{\omega} \tau} \hat{b}^\dagger ,
\end{eqnarray}
where $\hat{b}^\dagger (\tau)$ corresponds to the interaction representation of $\hat{b}^\dagger$ but it is not a Hermitian conjugate of $\hat{b}(\tau)$ because $\tau$ is imaginary time. 
Using Campbell-Baker-Hausdorff identity and the following thermodynamic average,
\begin{eqnarray}
	&& \langle e^{z_1 \hat{b}^\dagger} e^{z_2 \hat{b}} \rangle = \exp\left(z_1 z_2 N_{\mathrm{ph}}\right) , \label{identity}
\end{eqnarray}
where $z_1$, $z_2$ are arbitrary complex numbers, $N_{\mathrm{ph}}$ the average number of photons in the harmonic oscillator, see Eq.~(\ref{Nph}), we find the photon correlator,
\begin{eqnarray}
	&& \hspace{-26pt} D_{\nu_1, \nu_2} (\tau)  = \Phi_0^2 \nonumber \\
	&& \hspace{-23pt} \times \left(\exp\left[- \nu_1 \nu_2 \tilde{g}^2 \left(2 N_{\mathrm{ph}} \cosh(\tilde{\omega} \tau) + e^{-\tilde{\omega} |\tau|}\right)\right] - 1 \right) \! . \label{D}
\end{eqnarray}
As $\tilde{g} \ll 1$, we can expand this expression to the leading order,
\begin{eqnarray}
	&& \hspace{-15pt} D_{\nu_1, \nu_2} (\tau) \approx - \nu_1 \nu_2 \tilde{g}^2 \left(2 N_{\mathrm{ph}} \cosh(\tilde{\omega} \tau) + e^{-\tilde{\omega} |\tau|}\right). \label{Dexpand}
\end{eqnarray}
As $D_{\nu_1, \nu_2} (\tau)$ depends only on the product of indices $\nu_1 \nu_2$, we can simplify $\mathcal{F}_{\mathrm{int}}$,
\begin{eqnarray}
	&& \hspace{-29pt} \mathcal{F}_{\mathrm{int}} \! = \! - \frac{\Delta_0^2}{8} \! \int\limits_0^{\beta} \! D_{+,+} (\tau) \! \sum\limits_{\nu} \! \left[\Pi_{\nu, \nu} (\tau) - \Pi_{\nu, -\nu} (\tau)\right]  d\tau . \label{F1}
\end{eqnarray}

The dipole-dipole correlator $\Pi_{\nu_1, \nu_2}(\tau)$ can be represented in terms of the fermion Green's function $G_{\bm k} (\tau)$,
\begin{eqnarray}
	&& \hspace{-12pt} \Pi_{\nu_1, \nu_2}(\tau) = - \frac{1}{4} \sum\limits_{\bm k} \mathrm{Tr} \left\{\eta_{\nu_1} G_{\bm k} (\tau) \eta_{\nu_2} G_{\bm k} (-\tau) \right\} , \label{Pidiag} \\
	&& \hspace{-12pt} G_{\bm k}(\tau) = - \left\langle \mathcal{T} \left\{ \Psi_{\bm k} (\tau) \Psi_{\bm k}^\dagger (0)  \right\} \right\rangle \nonumber \\
	&& \hspace{17pt} = \sum\limits_{\sigma} |\sigma\rangle \langle \sigma| e^{-\xi_{\bm k, \sigma} \tau} \left(n_{\bm k, \sigma} - \vartheta(\tau) \right) ,
\end{eqnarray}
where we substituted Eq.~(\ref{Jnu}) into Eq.~(\ref{Pi}) and applied the Wick's theorem, the trace $\mathrm{Tr}$ is taken over the spin and the subband index, $\eta_\pm = \eta_x \pm i \eta_y$ are the Pauli matrices describing the dipole moment.
The electron Green's function is expanded with respect to the projectors on two bands characterized by the dipole spinors $|\pm\rangle$, see Eq.~(\ref{U}), $\xi_{\bm k, \sigma}$ is the renormalized spectrum corresponding to the band $\sigma$, see Eq.~(\ref{tildexik}), $n_{\bm k, \sigma} = [e^{\beta \xi_{\bm k, \sigma}} + 1]^{-1}$ is the Fermi-Dirac distribution function, $\vartheta(\tau) = 0$ ($\vartheta(\tau) = 1$) at $\tau < 0$ ($\tau > 0$) is the Heaviside step function.
The spin trace contributes the factor of $2$.
Evaluating the trace over the subband index, we find,
\begin{eqnarray}
	&& \hspace{-35pt} \sum\limits_{\nu} \left[\Pi_{\nu, \nu} (\tau) - \Pi_{\nu, -\nu} (\tau)\right] \nonumber \\
	&& = - 2 \sum\limits_{\bm k, \sigma} e^{\sigma E_g |\tau|} n_{\bm k, -\sigma} \left(1 - n_{\bm k, \sigma}\right) \nonumber \\
	&&  = -\frac{m \Omega E_g}{\pi} \frac{\cosh \left(E_g \left(|\tau| - \frac{\beta}{2}\right)\right)}{\sinh \left(\frac{\beta E_g}{2}\right)} S , \label{PiminusPi} \\
	&& \hspace{-35pt} S =  \left[1 - \frac{T}{E_g} \ln\left|\frac{1 + \exp\left(-\beta \left(E_F - \frac{E_g}{2}\right)\right)}{1 + \exp\left(-\beta \left(E_F + \frac{E_g}{2}\right)\right)}\right|\right]. \label{S1}
\end{eqnarray}
As we assume $T \ll E_F - E_g/2$, we can use the approximation $S \approx 1$.

Substituting Eqs.~(\ref{Dexpand}), (\ref{PiminusPi}) with $S = 1$ into Eq.~(\ref{F1}) and evaluating elementary integral over $\tau$, we obtain Eq.~(\ref{Fintexplicit}) in the main text.

\bibliography{bibliography1}

\begin{thebibliography}{51}%
\makeatletter
\providecommand \@ifxundefined [1]{%
 \@ifx{#1\undefined}
}%
\providecommand \@ifnum [1]{%
 \ifnum #1\expandafter \@firstoftwo
 \else \expandafter \@secondoftwo
 \fi
}%
\providecommand \@ifx [1]{%
 \ifx #1\expandafter \@firstoftwo
 \else \expandafter \@secondoftwo
 \fi
}%
\providecommand \natexlab [1]{#1}%
\providecommand \enquote  [1]{``#1''}%
\providecommand \bibnamefont  [1]{#1}%
\providecommand \bibfnamefont [1]{#1}%
\providecommand \citenamefont [1]{#1}%
\providecommand \href@noop [0]{\@secondoftwo}%
\providecommand \href [0]{\begingroup \@sanitize@url \@href}%
\providecommand \@href[1]{\@@startlink{#1}\@@href}%
\providecommand \@@href[1]{\endgroup#1\@@endlink}%
\providecommand \@sanitize@url [0]{\catcode `\\12\catcode `\$12\catcode
  `\&12\catcode `\#12\catcode `\^12\catcode `\_12\catcode `\%12\relax}%
\providecommand \@@startlink[1]{}%
\providecommand \@@endlink[0]{}%
\providecommand \url  [0]{\begingroup\@sanitize@url \@url }%
\providecommand \@url [1]{\endgroup\@href {#1}{\urlprefix }}%
\providecommand \urlprefix  [0]{URL }%
\providecommand \Eprint [0]{\href }%
\providecommand \doibase [0]{https://doi.org/}%
\providecommand \selectlanguage [0]{\@gobble}%
\providecommand \bibinfo  [0]{\@secondoftwo}%
\providecommand \bibfield  [0]{\@secondoftwo}%
\providecommand \translation [1]{[#1]}%
\providecommand \BibitemOpen [0]{}%
\providecommand \bibitemStop [0]{}%
\providecommand \bibitemNoStop [0]{.\EOS\space}%
\providecommand \EOS [0]{\spacefactor3000\relax}%
\providecommand \BibitemShut  [1]{\csname bibitem#1\endcsname}%
\let\auto@bib@innerbib\@empty
\bibitem [{\citenamefont {Schlawin}\ \emph {et~al.}(2022)\citenamefont
  {Schlawin}, \citenamefont {Kennes},\ and\ \citenamefont
  {Sentef}}]{cavQuantMatReview}%
  \BibitemOpen
  \bibfield  {author} {\bibinfo {author} {\bibfnamefont {F.}~\bibnamefont
  {Schlawin}}, \bibinfo {author} {\bibfnamefont {D.~M.}\ \bibnamefont
  {Kennes}},\ and\ \bibinfo {author} {\bibfnamefont {M.~A.}\ \bibnamefont
  {Sentef}},\ }\bibfield  {title} {\bibinfo {title} {Cavity quantum
  materials},\ }\href {https://doi.org/10.1063/5.0083825} {\bibfield  {journal}
  {\bibinfo  {journal} {Applied Physics Reviews}\ }\textbf {\bibinfo {volume}
  {9}},\ \bibinfo {pages} {011312} (\bibinfo {year} {2022})}\BibitemShut
  {NoStop}%
\bibitem [{\citenamefont {Qin}\ \emph {et~al.}(2024)\citenamefont {Qin},
  \citenamefont {Kockum}, \citenamefont {Muñoz}, \citenamefont {Miranowicz},\
  and\ \citenamefont {Nori}}]{QIN20241}%
  \BibitemOpen
  \bibfield  {author} {\bibinfo {author} {\bibfnamefont {W.}~\bibnamefont
  {Qin}}, \bibinfo {author} {\bibfnamefont {A.~F.}\ \bibnamefont {Kockum}},
  \bibinfo {author} {\bibfnamefont {C.~S.}\ \bibnamefont {Muñoz}}, \bibinfo
  {author} {\bibfnamefont {A.}~\bibnamefont {Miranowicz}},\ and\ \bibinfo
  {author} {\bibfnamefont {F.}~\bibnamefont {Nori}},\ }\bibfield  {title}
  {\bibinfo {title} {Quantum amplification and simulation of strong and
  ultrastrong coupling of light and matter},\ }\href
  {https://doi.org/https://doi.org/10.1016/j.physrep.2024.05.003} {\bibfield
  {journal} {\bibinfo  {journal} {Physics Reports}\ }\textbf {\bibinfo {volume}
  {1078}},\ \bibinfo {pages} {1} (\bibinfo {year} {2024})}\BibitemShut
  {NoStop}%
\bibitem [{\citenamefont {Kibis}(2010)}]{PhysRevB.81.165433}%
  \BibitemOpen
  \bibfield  {author} {\bibinfo {author} {\bibfnamefont {O.~V.}\ \bibnamefont
  {Kibis}},\ }\bibfield  {title} {\bibinfo {title} {Metal-insulator transition
  in graphene induced by circularly polarized photons},\ }\href
  {https://doi.org/10.1103/PhysRevB.81.165433} {\bibfield  {journal} {\bibinfo
  {journal} {Phys. Rev. B}\ }\textbf {\bibinfo {volume} {81}},\ \bibinfo
  {pages} {165433} (\bibinfo {year} {2010})}\BibitemShut {NoStop}%
\bibitem [{\citenamefont {Oka}\ and\ \citenamefont
  {Aoki}(2009)}]{PhysRevB.79.081406}%
  \BibitemOpen
  \bibfield  {author} {\bibinfo {author} {\bibfnamefont {T.}~\bibnamefont
  {Oka}}\ and\ \bibinfo {author} {\bibfnamefont {H.}~\bibnamefont {Aoki}},\
  }\bibfield  {title} {\bibinfo {title} {Photovoltaic hall effect in
  graphene},\ }\href {https://doi.org/10.1103/PhysRevB.79.081406} {\bibfield
  {journal} {\bibinfo  {journal} {Phys. Rev. B}\ }\textbf {\bibinfo {volume}
  {79}},\ \bibinfo {pages} {081406} (\bibinfo {year} {2009})}\BibitemShut
  {NoStop}%
\bibitem [{\citenamefont {Lindner}\ \emph {et~al.}(2011)\citenamefont
  {Lindner}, \citenamefont {Refael},\ and\ \citenamefont
  {Galitski}}]{Lindner2011}%
  \BibitemOpen
  \bibfield  {author} {\bibinfo {author} {\bibfnamefont {N.~H.}\ \bibnamefont
  {Lindner}}, \bibinfo {author} {\bibfnamefont {G.}~\bibnamefont {Refael}},\
  and\ \bibinfo {author} {\bibfnamefont {V.}~\bibnamefont {Galitski}},\
  }\bibfield  {title} {\bibinfo {title} {Floquet topological insulator in
  semiconductor quantum wells},\ }\href {https://doi.org/10.1038/nphys1926}
  {\bibfield  {journal} {\bibinfo  {journal} {Nature Physics}\ }\textbf
  {\bibinfo {volume} {7}},\ \bibinfo {pages} {490} (\bibinfo {year}
  {2011})}\BibitemShut {NoStop}%
\bibitem [{\citenamefont {Dehghani}\ \emph {et~al.}(2015)\citenamefont
  {Dehghani}, \citenamefont {Oka},\ and\ \citenamefont {Mitra}}]{Dehghani2015}%
  \BibitemOpen
  \bibfield  {author} {\bibinfo {author} {\bibfnamefont {H.}~\bibnamefont
  {Dehghani}}, \bibinfo {author} {\bibfnamefont {T.}~\bibnamefont {Oka}},\ and\
  \bibinfo {author} {\bibfnamefont {A.}~\bibnamefont {Mitra}},\ }\bibfield
  {title} {\bibinfo {title} {Out-of-equilibrium electrons and the hall
  conductance of a floquet topological insulator},\ }\href
  {https://doi.org/10.1103/PhysRevB.91.155422} {\bibfield  {journal} {\bibinfo
  {journal} {Phys. Rev. B}\ }\textbf {\bibinfo {volume} {91}},\ \bibinfo
  {pages} {155422} (\bibinfo {year} {2015})}\BibitemShut {NoStop}%
\bibitem [{\citenamefont {Thakurathi}\ \emph {et~al.}(2017)\citenamefont
  {Thakurathi}, \citenamefont {Loss},\ and\ \citenamefont
  {Klinovaja}}]{PhysRevB.95.155407}%
  \BibitemOpen
  \bibfield  {author} {\bibinfo {author} {\bibfnamefont {M.}~\bibnamefont
  {Thakurathi}}, \bibinfo {author} {\bibfnamefont {D.}~\bibnamefont {Loss}},\
  and\ \bibinfo {author} {\bibfnamefont {J.}~\bibnamefont {Klinovaja}},\
  }\bibfield  {title} {\bibinfo {title} {Floquet majorana fermions and
  parafermions in driven rashba nanowires},\ }\href
  {https://doi.org/10.1103/PhysRevB.95.155407} {\bibfield  {journal} {\bibinfo
  {journal} {Phys. Rev. B}\ }\textbf {\bibinfo {volume} {95}},\ \bibinfo
  {pages} {155407} (\bibinfo {year} {2017})}\BibitemShut {NoStop}%
\bibitem [{\citenamefont {Klinovaja}\ \emph {et~al.}(2016)\citenamefont
  {Klinovaja}, \citenamefont {Stano},\ and\ \citenamefont
  {Loss}}]{PhysRevLett.116.176401}%
  \BibitemOpen
  \bibfield  {author} {\bibinfo {author} {\bibfnamefont {J.}~\bibnamefont
  {Klinovaja}}, \bibinfo {author} {\bibfnamefont {P.}~\bibnamefont {Stano}},\
  and\ \bibinfo {author} {\bibfnamefont {D.}~\bibnamefont {Loss}},\ }\bibfield
  {title} {\bibinfo {title} {Topological floquet phases in driven coupled
  rashba nanowires},\ }\href {https://doi.org/10.1103/PhysRevLett.116.176401}
  {\bibfield  {journal} {\bibinfo  {journal} {Phys. Rev. Lett.}\ }\textbf
  {\bibinfo {volume} {116}},\ \bibinfo {pages} {176401} (\bibinfo {year}
  {2016})}\BibitemShut {NoStop}%
\bibitem [{\citenamefont {Thakurathi}\ \emph {et~al.}(2013)\citenamefont
  {Thakurathi}, \citenamefont {Patel}, \citenamefont {Sen},\ and\ \citenamefont
  {Dutta}}]{PhysRevB.88.155133}%
  \BibitemOpen
  \bibfield  {author} {\bibinfo {author} {\bibfnamefont {M.}~\bibnamefont
  {Thakurathi}}, \bibinfo {author} {\bibfnamefont {A.~A.}\ \bibnamefont
  {Patel}}, \bibinfo {author} {\bibfnamefont {D.}~\bibnamefont {Sen}},\ and\
  \bibinfo {author} {\bibfnamefont {A.}~\bibnamefont {Dutta}},\ }\bibfield
  {title} {\bibinfo {title} {Floquet generation of majorana end modes and
  topological invariants},\ }\href {https://doi.org/10.1103/PhysRevB.88.155133}
  {\bibfield  {journal} {\bibinfo  {journal} {Phys. Rev. B}\ }\textbf {\bibinfo
  {volume} {88}},\ \bibinfo {pages} {155133} (\bibinfo {year}
  {2013})}\BibitemShut {NoStop}%
\bibitem [{\citenamefont {Kundu}\ and\ \citenamefont
  {Seradjeh}(2013)}]{PhysRevLett.111.136402}%
  \BibitemOpen
  \bibfield  {author} {\bibinfo {author} {\bibfnamefont {A.}~\bibnamefont
  {Kundu}}\ and\ \bibinfo {author} {\bibfnamefont {B.}~\bibnamefont
  {Seradjeh}},\ }\bibfield  {title} {\bibinfo {title} {Transport signatures of
  floquet majorana fermions in driven topological superconductors},\ }\href
  {https://doi.org/10.1103/PhysRevLett.111.136402} {\bibfield  {journal}
  {\bibinfo  {journal} {Phys. Rev. Lett.}\ }\textbf {\bibinfo {volume} {111}},\
  \bibinfo {pages} {136402} (\bibinfo {year} {2013})}\BibitemShut {NoStop}%
\bibitem [{\citenamefont {Ricco}\ \emph
  {et~al.}(2022{\natexlab{a}})\citenamefont {Ricco}, \citenamefont {Kozin},
  \citenamefont {Seridonio},\ and\ \citenamefont {Shelykh}}]{Ricco2022}%
  \BibitemOpen
  \bibfield  {author} {\bibinfo {author} {\bibfnamefont {L.~S.}\ \bibnamefont
  {Ricco}}, \bibinfo {author} {\bibfnamefont {V.~K.}\ \bibnamefont {Kozin}},
  \bibinfo {author} {\bibfnamefont {A.~C.}\ \bibnamefont {Seridonio}},\ and\
  \bibinfo {author} {\bibfnamefont {I.~A.}\ \bibnamefont {Shelykh}},\
  }\bibfield  {title} {\bibinfo {title} {Accessing the degree of majorana
  nonlocality in a quantum dot-optical microcavity system},\ }\href
  {https://doi.org/10.1038/s41598-022-05855-y} {\bibfield  {journal} {\bibinfo
  {journal} {Scientific Reports}\ }\textbf {\bibinfo {volume} {12}},\ \bibinfo
  {pages} {1983} (\bibinfo {year} {2022}{\natexlab{a}})}\BibitemShut {NoStop}%
\bibitem [{\citenamefont {Ricco}\ \emph
  {et~al.}(2022{\natexlab{b}})\citenamefont {Ricco}, \citenamefont {Kozin},
  \citenamefont {Seridonio},\ and\ \citenamefont
  {Shelykh}}]{PhysRevA.106.023702}%
  \BibitemOpen
  \bibfield  {author} {\bibinfo {author} {\bibfnamefont {L.~S.}\ \bibnamefont
  {Ricco}}, \bibinfo {author} {\bibfnamefont {V.~K.}\ \bibnamefont {Kozin}},
  \bibinfo {author} {\bibfnamefont {A.~C.}\ \bibnamefont {Seridonio}},\ and\
  \bibinfo {author} {\bibfnamefont {I.~A.}\ \bibnamefont {Shelykh}},\
  }\bibfield  {title} {\bibinfo {title} {Reshaping the jaynes-cummings ladder
  with majorana bound states},\ }\href
  {https://doi.org/10.1103/PhysRevA.106.023702} {\bibfield  {journal} {\bibinfo
   {journal} {Phys. Rev. A}\ }\textbf {\bibinfo {volume} {106}},\ \bibinfo
  {pages} {023702} (\bibinfo {year} {2022}{\natexlab{b}})}\BibitemShut
  {NoStop}%
\bibitem [{\citenamefont {Wang}\ \emph {et~al.}(2013)\citenamefont {Wang},
  \citenamefont {Steinberg}, \citenamefont {Jarillo-Herrero},\ and\
  \citenamefont {Gedik}}]{doi:10.1126/science.1239834}%
  \BibitemOpen
  \bibfield  {author} {\bibinfo {author} {\bibfnamefont {Y.~H.}\ \bibnamefont
  {Wang}}, \bibinfo {author} {\bibfnamefont {H.}~\bibnamefont {Steinberg}},
  \bibinfo {author} {\bibfnamefont {P.}~\bibnamefont {Jarillo-Herrero}},\ and\
  \bibinfo {author} {\bibfnamefont {N.}~\bibnamefont {Gedik}},\ }\bibfield
  {title} {\bibinfo {title} {Observation of floquet-bloch states on the surface
  of a topological insulator},\ }\href
  {https://doi.org/10.1126/science.1239834} {\bibfield  {journal} {\bibinfo
  {journal} {Science}\ }\textbf {\bibinfo {volume} {342}},\ \bibinfo {pages}
  {453} (\bibinfo {year} {2013})}\BibitemShut {NoStop}%
\bibitem [{\citenamefont {McIver}\ \emph {et~al.}(2020)\citenamefont {McIver},
  \citenamefont {Schulte}, \citenamefont {Stein}, \citenamefont {Matsuyama},
  \citenamefont {Jotzu}, \citenamefont {Meier},\ and\ \citenamefont
  {Cavalleri}}]{McIver2020}%
  \BibitemOpen
  \bibfield  {author} {\bibinfo {author} {\bibfnamefont {J.~W.}\ \bibnamefont
  {McIver}}, \bibinfo {author} {\bibfnamefont {B.}~\bibnamefont {Schulte}},
  \bibinfo {author} {\bibfnamefont {F.-U.}\ \bibnamefont {Stein}}, \bibinfo
  {author} {\bibfnamefont {T.}~\bibnamefont {Matsuyama}}, \bibinfo {author}
  {\bibfnamefont {G.}~\bibnamefont {Jotzu}}, \bibinfo {author} {\bibfnamefont
  {G.}~\bibnamefont {Meier}},\ and\ \bibinfo {author} {\bibfnamefont
  {A.}~\bibnamefont {Cavalleri}},\ }\bibfield  {title} {\bibinfo {title}
  {Light-induced anomalous hall effect in graphene},\ }\href
  {https://doi.org/10.1038/s41567-019-0698-y} {\bibfield  {journal} {\bibinfo
  {journal} {Nature Physics}\ }\textbf {\bibinfo {volume} {16}},\ \bibinfo
  {pages} {38} (\bibinfo {year} {2020})}\BibitemShut {NoStop}%
\bibitem [{\citenamefont {Kozin}\ \emph
  {et~al.}(2018{\natexlab{a}})\citenamefont {Kozin}, \citenamefont {Iorsh},
  \citenamefont {Kibis},\ and\ \citenamefont {Shelykh}}]{PhysRevB.97.035416}%
  \BibitemOpen
  \bibfield  {author} {\bibinfo {author} {\bibfnamefont {V.~K.}\ \bibnamefont
  {Kozin}}, \bibinfo {author} {\bibfnamefont {I.~V.}\ \bibnamefont {Iorsh}},
  \bibinfo {author} {\bibfnamefont {O.~V.}\ \bibnamefont {Kibis}},\ and\
  \bibinfo {author} {\bibfnamefont {I.~A.}\ \bibnamefont {Shelykh}},\
  }\bibfield  {title} {\bibinfo {title} {Periodic array of quantum rings
  strongly coupled to circularly polarized light as a topological insulator},\
  }\href {https://doi.org/10.1103/PhysRevB.97.035416} {\bibfield  {journal}
  {\bibinfo  {journal} {Phys. Rev. B}\ }\textbf {\bibinfo {volume} {97}},\
  \bibinfo {pages} {035416} (\bibinfo {year} {2018}{\natexlab{a}})}\BibitemShut
  {NoStop}%
\bibitem [{\citenamefont {Cooper}\ \emph {et~al.}(2019)\citenamefont {Cooper},
  \citenamefont {Dalibard},\ and\ \citenamefont
  {Spielman}}]{RevModPhys.91.015005}%
  \BibitemOpen
  \bibfield  {author} {\bibinfo {author} {\bibfnamefont {N.~R.}\ \bibnamefont
  {Cooper}}, \bibinfo {author} {\bibfnamefont {J.}~\bibnamefont {Dalibard}},\
  and\ \bibinfo {author} {\bibfnamefont {I.~B.}\ \bibnamefont {Spielman}},\
  }\bibfield  {title} {\bibinfo {title} {Topological bands for ultracold
  atoms},\ }\href {https://doi.org/10.1103/RevModPhys.91.015005} {\bibfield
  {journal} {\bibinfo  {journal} {Rev. Mod. Phys.}\ }\textbf {\bibinfo {volume}
  {91}},\ \bibinfo {pages} {015005} (\bibinfo {year} {2019})}\BibitemShut
  {NoStop}%
\bibitem [{\citenamefont {Roux}\ \emph {et~al.}(2020)\citenamefont {Roux},
  \citenamefont {Konishi}, \citenamefont {Helson},\ and\ \citenamefont
  {Brantut}}]{Roux2020}%
  \BibitemOpen
  \bibfield  {author} {\bibinfo {author} {\bibfnamefont {K.}~\bibnamefont
  {Roux}}, \bibinfo {author} {\bibfnamefont {H.}~\bibnamefont {Konishi}},
  \bibinfo {author} {\bibfnamefont {V.}~\bibnamefont {Helson}},\ and\ \bibinfo
  {author} {\bibfnamefont {J.-P.}\ \bibnamefont {Brantut}},\ }\bibfield
  {title} {\bibinfo {title} {Strongly correlated fermions strongly coupled to
  light},\ }\href {https://doi.org/10.1038/s41467-020-16767-8} {\bibfield
  {journal} {\bibinfo  {journal} {Nature Communications}\ }\textbf {\bibinfo
  {volume} {11}},\ \bibinfo {pages} {2974} (\bibinfo {year}
  {2020})}\BibitemShut {NoStop}%
\bibitem [{\citenamefont {Mivehvar}\ \emph {et~al.}(2017)\citenamefont
  {Mivehvar}, \citenamefont {Ritsch},\ and\ \citenamefont
  {Piazza}}]{PhysRevLett.118.073602}%
  \BibitemOpen
  \bibfield  {author} {\bibinfo {author} {\bibfnamefont {F.}~\bibnamefont
  {Mivehvar}}, \bibinfo {author} {\bibfnamefont {H.}~\bibnamefont {Ritsch}},\
  and\ \bibinfo {author} {\bibfnamefont {F.}~\bibnamefont {Piazza}},\
  }\bibfield  {title} {\bibinfo {title} {Superradiant topological peierls
  insulator inside an optical cavity},\ }\href
  {https://doi.org/10.1103/PhysRevLett.118.073602} {\bibfield  {journal}
  {\bibinfo  {journal} {Phys. Rev. Lett.}\ }\textbf {\bibinfo {volume} {118}},\
  \bibinfo {pages} {073602} (\bibinfo {year} {2017})}\BibitemShut {NoStop}%
\bibitem [{\citenamefont {Sedov}\ \emph {et~al.}(2020)\citenamefont {Sedov},
  \citenamefont {Kozin},\ and\ \citenamefont {Iorsh}}]{PhysRevLett.125.263606}%
  \BibitemOpen
  \bibfield  {author} {\bibinfo {author} {\bibfnamefont {D.~D.}\ \bibnamefont
  {Sedov}}, \bibinfo {author} {\bibfnamefont {V.~K.}\ \bibnamefont {Kozin}},\
  and\ \bibinfo {author} {\bibfnamefont {I.~V.}\ \bibnamefont {Iorsh}},\
  }\bibfield  {title} {\bibinfo {title} {Chiral waveguide optomechanics: First
  order quantum phase transitions with ${\mathbb{z}}_{3}$ symmetry breaking},\
  }\href {https://doi.org/10.1103/PhysRevLett.125.263606} {\bibfield  {journal}
  {\bibinfo  {journal} {Phys. Rev. Lett.}\ }\textbf {\bibinfo {volume} {125}},\
  \bibinfo {pages} {263606} (\bibinfo {year} {2020})}\BibitemShut {NoStop}%
\bibitem [{\citenamefont {Kozin}\ \emph {et~al.}(2025)\citenamefont {Kozin},
  \citenamefont {Thingstad}, \citenamefont {Loss},\ and\ \citenamefont
  {Klinovaja}}]{kozin2024cavityenhancedsuperconductivitybandengineering}%
  \BibitemOpen
  \bibfield  {author} {\bibinfo {author} {\bibfnamefont {V.~K.}\ \bibnamefont
  {Kozin}}, \bibinfo {author} {\bibfnamefont {E.}~\bibnamefont {Thingstad}},
  \bibinfo {author} {\bibfnamefont {D.}~\bibnamefont {Loss}},\ and\ \bibinfo
  {author} {\bibfnamefont {J.}~\bibnamefont {Klinovaja}},\ }\bibfield  {title}
  {\bibinfo {title} {Cavity-enhanced superconductivity via band engineering},\
  }\href {https://doi.org/10.1103/PhysRevB.111.035410} {\bibfield  {journal}
  {\bibinfo  {journal} {Phys. Rev. B}\ }\textbf {\bibinfo {volume} {111}},\
  \bibinfo {pages} {035410} (\bibinfo {year} {2025})}\BibitemShut {NoStop}%
\bibitem [{\citenamefont {Sentef}\ \emph {et~al.}(2018)\citenamefont {Sentef},
  \citenamefont {Ruggenthaler},\ and\ \citenamefont {Rubio}}]{Sentef2018}%
  \BibitemOpen
  \bibfield  {author} {\bibinfo {author} {\bibfnamefont {M.~A.}\ \bibnamefont
  {Sentef}}, \bibinfo {author} {\bibfnamefont {M.}~\bibnamefont
  {Ruggenthaler}},\ and\ \bibinfo {author} {\bibfnamefont {A.}~\bibnamefont
  {Rubio}},\ }\bibfield  {title} {\bibinfo {title} {Cavity
  quantum-electrodynamical polaritonically enhanced electron-phonon coupling
  and its influence on superconductivity},\ }\bibfield  {journal} {\bibinfo
  {journal} {Science Advances}\ }\textbf {\bibinfo {volume} {4}},\ \href
  {https://doi.org/10.1126/sciadv.aau6969} {10.1126/sciadv.aau6969} (\bibinfo
  {year} {2018})\BibitemShut {NoStop}%
\bibitem [{\citenamefont {Li}\ and\ \citenamefont
  {Eckstein}(2020)}]{PhysRevLett.125.217402}%
  \BibitemOpen
  \bibfield  {author} {\bibinfo {author} {\bibfnamefont {J.}~\bibnamefont
  {Li}}\ and\ \bibinfo {author} {\bibfnamefont {M.}~\bibnamefont {Eckstein}},\
  }\bibfield  {title} {\bibinfo {title} {Manipulating intertwined orders in
  solids with quantum light},\ }\href
  {https://doi.org/10.1103/PhysRevLett.125.217402} {\bibfield  {journal}
  {\bibinfo  {journal} {Phys. Rev. Lett.}\ }\textbf {\bibinfo {volume} {125}},\
  \bibinfo {pages} {217402} (\bibinfo {year} {2020})}\BibitemShut {NoStop}%
\bibitem [{\citenamefont {Chakraborty}\ and\ \citenamefont
  {Piazza}(2021)}]{DrivingFluctuationsPairing}%
  \BibitemOpen
  \bibfield  {author} {\bibinfo {author} {\bibfnamefont {A.}~\bibnamefont
  {Chakraborty}}\ and\ \bibinfo {author} {\bibfnamefont {F.}~\bibnamefont
  {Piazza}},\ }\bibfield  {title} {\bibinfo {title} {Long-range photon
  fluctuations enhance photon-mediated electron pairing and
  superconductivity},\ }\href {https://doi.org/10.1103/PhysRevLett.127.177002}
  {\bibfield  {journal} {\bibinfo  {journal} {Phys. Rev. Lett.}\ }\textbf
  {\bibinfo {volume} {127}},\ \bibinfo {pages} {177002} (\bibinfo {year}
  {2021})}\BibitemShut {NoStop}%
\bibitem [{\citenamefont {Gao}\ \emph {et~al.}(2020)\citenamefont {Gao},
  \citenamefont {Schlawin}, \citenamefont {Buzzi}, \citenamefont {Cavalleri},\
  and\ \citenamefont {Jaksch}}]{DrivingPairing}%
  \BibitemOpen
  \bibfield  {author} {\bibinfo {author} {\bibfnamefont {H.}~\bibnamefont
  {Gao}}, \bibinfo {author} {\bibfnamefont {F.}~\bibnamefont {Schlawin}},
  \bibinfo {author} {\bibfnamefont {M.}~\bibnamefont {Buzzi}}, \bibinfo
  {author} {\bibfnamefont {A.}~\bibnamefont {Cavalleri}},\ and\ \bibinfo
  {author} {\bibfnamefont {D.}~\bibnamefont {Jaksch}},\ }\bibfield  {title}
  {\bibinfo {title} {Photoinduced electron pairing in a driven cavity},\ }\href
  {https://doi.org/10.1103/PhysRevLett.125.053602} {\bibfield  {journal}
  {\bibinfo  {journal} {Phys. Rev. Lett.}\ }\textbf {\bibinfo {volume} {125}},\
  \bibinfo {pages} {053602} (\bibinfo {year} {2020})}\BibitemShut {NoStop}%
\bibitem [{\citenamefont {Gao}\ \emph {et~al.}(2021)\citenamefont {Gao},
  \citenamefont {Schlawin},\ and\ \citenamefont
  {Jaksch}}]{PhysRevB.104.L140503}%
  \BibitemOpen
  \bibfield  {author} {\bibinfo {author} {\bibfnamefont {H.}~\bibnamefont
  {Gao}}, \bibinfo {author} {\bibfnamefont {F.}~\bibnamefont {Schlawin}},\ and\
  \bibinfo {author} {\bibfnamefont {D.}~\bibnamefont {Jaksch}},\ }\bibfield
  {title} {\bibinfo {title} {Higgs mode stabilization by photoinduced
  long-range interactions in a superconductor},\ }\href
  {https://doi.org/10.1103/PhysRevB.104.L140503} {\bibfield  {journal}
  {\bibinfo  {journal} {Phys. Rev. B}\ }\textbf {\bibinfo {volume} {104}},\
  \bibinfo {pages} {L140503} (\bibinfo {year} {2021})}\BibitemShut {NoStop}%
\bibitem [{\citenamefont {Eckhardt}\ \emph {et~al.}(2023)\citenamefont
  {Eckhardt}, \citenamefont {Chattopadhyay}, \citenamefont {Kennes},
  \citenamefont {Demler}, \citenamefont {Sentef},\ and\ \citenamefont
  {Michael}}]{eckhardt2023theory}%
  \BibitemOpen
  \bibfield  {author} {\bibinfo {author} {\bibfnamefont {C.~J.}\ \bibnamefont
  {Eckhardt}}, \bibinfo {author} {\bibfnamefont {S.}~\bibnamefont
  {Chattopadhyay}}, \bibinfo {author} {\bibfnamefont {D.~M.}\ \bibnamefont
  {Kennes}}, \bibinfo {author} {\bibfnamefont {E.~A.}\ \bibnamefont {Demler}},
  \bibinfo {author} {\bibfnamefont {M.~A.}\ \bibnamefont {Sentef}},\ and\
  \bibinfo {author} {\bibfnamefont {M.~H.}\ \bibnamefont {Michael}},\
  }\href@noop {} {} (\bibinfo {year} {2023}),\ \Eprint
  {https://arxiv.org/abs/2303.02176} {arXiv:2303.02176} \BibitemShut {NoStop}%
\bibitem [{\citenamefont {Laussy}\ \emph {et~al.}(2010)\citenamefont {Laussy},
  \citenamefont {Kavokin},\ and\ \citenamefont
  {Shelykh}}]{PhysRevLett.104.106402}%
  \BibitemOpen
  \bibfield  {author} {\bibinfo {author} {\bibfnamefont {F.~P.}\ \bibnamefont
  {Laussy}}, \bibinfo {author} {\bibfnamefont {A.~V.}\ \bibnamefont
  {Kavokin}},\ and\ \bibinfo {author} {\bibfnamefont {I.~A.}\ \bibnamefont
  {Shelykh}},\ }\bibfield  {title} {\bibinfo {title} {Exciton-polariton
  mediated superconductivity},\ }\href
  {https://doi.org/10.1103/PhysRevLett.104.106402} {\bibfield  {journal}
  {\bibinfo  {journal} {Phys. Rev. Lett.}\ }\textbf {\bibinfo {volume} {104}},\
  \bibinfo {pages} {106402} (\bibinfo {year} {2010})}\BibitemShut {NoStop}%
\bibitem [{\citenamefont {Kockum}\ \emph {et~al.}(2019)\citenamefont {Kockum},
  \citenamefont {Miranowicz}, \citenamefont {Liberato}, \citenamefont
  {Savasta},\ and\ \citenamefont {Nori}}]{FriskKockum2019}%
  \BibitemOpen
  \bibfield  {author} {\bibinfo {author} {\bibfnamefont {A.~F.}\ \bibnamefont
  {Kockum}}, \bibinfo {author} {\bibfnamefont {A.}~\bibnamefont {Miranowicz}},
  \bibinfo {author} {\bibfnamefont {S.~D.}\ \bibnamefont {Liberato}}, \bibinfo
  {author} {\bibfnamefont {S.}~\bibnamefont {Savasta}},\ and\ \bibinfo {author}
  {\bibfnamefont {F.}~\bibnamefont {Nori}},\ }\bibfield  {title} {\bibinfo
  {title} {Ultrastrong coupling between light and matter},\ }\href
  {https://doi.org/10.1038/s42254-018-0006-2} {\bibfield  {journal} {\bibinfo
  {journal} {Nature Reviews Physics}\ }\textbf {\bibinfo {volume} {1}},\
  \bibinfo {pages} {19} (\bibinfo {year} {2019})}\BibitemShut {NoStop}%
\bibitem [{\citenamefont {Oka}\ and\ \citenamefont
  {Kitamura}(2019)}]{FloquetReview}%
  \BibitemOpen
  \bibfield  {author} {\bibinfo {author} {\bibfnamefont {T.}~\bibnamefont
  {Oka}}\ and\ \bibinfo {author} {\bibfnamefont {S.}~\bibnamefont {Kitamura}},\
  }\bibfield  {title} {\bibinfo {title} {Floquet engineering of quantum
  materials},\ }\href
  {https://doi.org/10.1146/annurev-conmatphys-031218-013423} {\bibfield
  {journal} {\bibinfo  {journal} {Annual Review of Condensed Matter Physics}\
  }\textbf {\bibinfo {volume} {10}},\ \bibinfo {pages} {387} (\bibinfo {year}
  {2019})}\BibitemShut {NoStop}%
\bibitem [{\citenamefont {Kavokin}\ \emph {et~al.}(2017)\citenamefont
  {Kavokin}, \citenamefont {Baumberg}, \citenamefont {Laussy},\ and\
  \citenamefont {Malpuech}}]{MicrocavitiesKavokin}%
  \BibitemOpen
  \bibfield  {author} {\bibinfo {author} {\bibfnamefont {A.}~\bibnamefont
  {Kavokin}}, \bibinfo {author} {\bibfnamefont {J.~J.}\ \bibnamefont
  {Baumberg}}, \bibinfo {author} {\bibfnamefont {F.~P.}\ \bibnamefont
  {Laussy}},\ and\ \bibinfo {author} {\bibfnamefont {G.}~\bibnamefont
  {Malpuech}},\ }\href@noop {} {\emph {\bibinfo {title} {Microcavities}}}\
  (\bibinfo  {publisher} {Oxford University Press},\ \bibinfo {year}
  {2017})\BibitemShut {NoStop}%
\bibitem [{\citenamefont {Kozin}\ \emph
  {et~al.}(2018{\natexlab{b}})\citenamefont {Kozin}, \citenamefont {Shelykh},
  \citenamefont {Nalitov},\ and\ \citenamefont {Iorsh}}]{PhysRevB.98.125115}%
  \BibitemOpen
  \bibfield  {author} {\bibinfo {author} {\bibfnamefont {V.~K.}\ \bibnamefont
  {Kozin}}, \bibinfo {author} {\bibfnamefont {I.~A.}\ \bibnamefont {Shelykh}},
  \bibinfo {author} {\bibfnamefont {A.~V.}\ \bibnamefont {Nalitov}},\ and\
  \bibinfo {author} {\bibfnamefont {I.~V.}\ \bibnamefont {Iorsh}},\ }\bibfield
  {title} {\bibinfo {title} {Topological metamaterials based on polariton
  rings},\ }\href {https://doi.org/10.1103/PhysRevB.98.125115} {\bibfield
  {journal} {\bibinfo  {journal} {Phys. Rev. B}\ }\textbf {\bibinfo {volume}
  {98}},\ \bibinfo {pages} {125115} (\bibinfo {year}
  {2018}{\natexlab{b}})}\BibitemShut {NoStop}%
\bibitem [{\citenamefont {Maissen}\ \emph {et~al.}(2014)\citenamefont
  {Maissen}, \citenamefont {Scalari}, \citenamefont {Valmorra}, \citenamefont
  {Beck}, \citenamefont {Faist}, \citenamefont {Cibella}, \citenamefont
  {Leoni}, \citenamefont {Reichl}, \citenamefont {Charpentier},\ and\
  \citenamefont {Wegscheider}}]{ModeCompressionPRB}%
  \BibitemOpen
  \bibfield  {author} {\bibinfo {author} {\bibfnamefont {C.}~\bibnamefont
  {Maissen}}, \bibinfo {author} {\bibfnamefont {G.}~\bibnamefont {Scalari}},
  \bibinfo {author} {\bibfnamefont {F.}~\bibnamefont {Valmorra}}, \bibinfo
  {author} {\bibfnamefont {M.}~\bibnamefont {Beck}}, \bibinfo {author}
  {\bibfnamefont {J.}~\bibnamefont {Faist}}, \bibinfo {author} {\bibfnamefont
  {S.}~\bibnamefont {Cibella}}, \bibinfo {author} {\bibfnamefont
  {R.}~\bibnamefont {Leoni}}, \bibinfo {author} {\bibfnamefont
  {C.}~\bibnamefont {Reichl}}, \bibinfo {author} {\bibfnamefont
  {C.}~\bibnamefont {Charpentier}},\ and\ \bibinfo {author} {\bibfnamefont
  {W.}~\bibnamefont {Wegscheider}},\ }\bibfield  {title} {\bibinfo {title}
  {Ultrastrong coupling in the near field of complementary split-ring
  resonators},\ }\href {https://doi.org/10.1103/PhysRevB.90.205309} {\bibfield
  {journal} {\bibinfo  {journal} {Phys. Rev. B}\ }\textbf {\bibinfo {volume}
  {90}},\ \bibinfo {pages} {205309} (\bibinfo {year} {2014})}\BibitemShut
  {NoStop}%
\bibitem [{\citenamefont {Keller}\ \emph {et~al.}(2017)\citenamefont {Keller},
  \citenamefont {Scalari}, \citenamefont {Cibella}, \citenamefont {Maissen},
  \citenamefont {Appugliese}, \citenamefont {Giovine}, \citenamefont {Leoni},
  \citenamefont {Beck},\ and\ \citenamefont {Faist}}]{ModeCompressionNanoLett}%
  \BibitemOpen
  \bibfield  {author} {\bibinfo {author} {\bibfnamefont {J.}~\bibnamefont
  {Keller}}, \bibinfo {author} {\bibfnamefont {G.}~\bibnamefont {Scalari}},
  \bibinfo {author} {\bibfnamefont {S.}~\bibnamefont {Cibella}}, \bibinfo
  {author} {\bibfnamefont {C.}~\bibnamefont {Maissen}}, \bibinfo {author}
  {\bibfnamefont {F.}~\bibnamefont {Appugliese}}, \bibinfo {author}
  {\bibfnamefont {E.}~\bibnamefont {Giovine}}, \bibinfo {author} {\bibfnamefont
  {R.}~\bibnamefont {Leoni}}, \bibinfo {author} {\bibfnamefont
  {M.}~\bibnamefont {Beck}},\ and\ \bibinfo {author} {\bibfnamefont
  {J.}~\bibnamefont {Faist}},\ }\bibfield  {title} {\bibinfo {title}
  {Few-electron ultrastrong light-matter coupling at 300 ghz with nanogap
  hybrid lc microcavities},\ }\href
  {https://doi.org/10.1021/acs.nanolett.7b03228} {\bibfield  {journal}
  {\bibinfo  {journal} {Nano Letters}\ }\textbf {\bibinfo {volume} {17}},\
  \bibinfo {pages} {7410} (\bibinfo {year} {2017})}\BibitemShut {NoStop}%
\bibitem [{\citenamefont {Appugliese}\ \emph {et~al.}(2022)\citenamefont
  {Appugliese}, \citenamefont {Enkner}, \citenamefont {Paravicini-Bagliani},
  \citenamefont {Beck}, \citenamefont {Reichl}, \citenamefont {Wegscheider},
  \citenamefont {Scalari}, \citenamefont {Ciuti},\ and\ \citenamefont
  {Faist}}]{FaistHallBreakdown}%
  \BibitemOpen
  \bibfield  {author} {\bibinfo {author} {\bibfnamefont {F.}~\bibnamefont
  {Appugliese}}, \bibinfo {author} {\bibfnamefont {J.}~\bibnamefont {Enkner}},
  \bibinfo {author} {\bibfnamefont {G.~L.}\ \bibnamefont
  {Paravicini-Bagliani}}, \bibinfo {author} {\bibfnamefont {M.}~\bibnamefont
  {Beck}}, \bibinfo {author} {\bibfnamefont {C.}~\bibnamefont {Reichl}},
  \bibinfo {author} {\bibfnamefont {W.}~\bibnamefont {Wegscheider}}, \bibinfo
  {author} {\bibfnamefont {G.}~\bibnamefont {Scalari}}, \bibinfo {author}
  {\bibfnamefont {C.}~\bibnamefont {Ciuti}},\ and\ \bibinfo {author}
  {\bibfnamefont {J.}~\bibnamefont {Faist}},\ }\bibfield  {title} {\bibinfo
  {title} {Breakdown of topological protection by cavity vacuum fields in the
  integer quantum hall effect},\ }\href
  {https://doi.org/10.1126/science.abl5818} {\bibfield  {journal} {\bibinfo
  {journal} {Science}\ }\textbf {\bibinfo {volume} {375}},\ \bibinfo {pages}
  {1030} (\bibinfo {year} {2022})}\BibitemShut {NoStop}%
\bibitem [{\citenamefont {Wang}\ \emph {et~al.}(2019)\citenamefont {Wang},
  \citenamefont {Ronca},\ and\ \citenamefont {Sentef}}]{PhysRevB.99.235156}%
  \BibitemOpen
  \bibfield  {author} {\bibinfo {author} {\bibfnamefont {X.}~\bibnamefont
  {Wang}}, \bibinfo {author} {\bibfnamefont {E.}~\bibnamefont {Ronca}},\ and\
  \bibinfo {author} {\bibfnamefont {M.~A.}\ \bibnamefont {Sentef}},\ }\bibfield
   {title} {\bibinfo {title} {Cavity quantum electrodynamical chern insulator:
  Towards light-induced quantized anomalous hall effect in graphene},\ }\href
  {https://doi.org/10.1103/PhysRevB.99.235156} {\bibfield  {journal} {\bibinfo
  {journal} {Phys. Rev. B}\ }\textbf {\bibinfo {volume} {99}},\ \bibinfo
  {pages} {235156} (\bibinfo {year} {2019})}\BibitemShut {NoStop}%
\bibitem [{\citenamefont {Kozin}\ \emph {et~al.}(2024)\citenamefont {Kozin},
  \citenamefont {Miserev}, \citenamefont {Loss},\ and\ \citenamefont
  {Klinovaja}}]{PhysRevResearch.6.033188}%
  \BibitemOpen
  \bibfield  {author} {\bibinfo {author} {\bibfnamefont {V.~K.}\ \bibnamefont
  {Kozin}}, \bibinfo {author} {\bibfnamefont {D.}~\bibnamefont {Miserev}},
  \bibinfo {author} {\bibfnamefont {D.}~\bibnamefont {Loss}},\ and\ \bibinfo
  {author} {\bibfnamefont {J.}~\bibnamefont {Klinovaja}},\ }\bibfield  {title}
  {\bibinfo {title} {Quantum phase transitions and cat states in cavity-coupled
  quantum dots},\ }\href {https://doi.org/10.1103/PhysRevResearch.6.033188}
  {\bibfield  {journal} {\bibinfo  {journal} {Phys. Rev. Res.}\ }\textbf
  {\bibinfo {volume} {6}},\ \bibinfo {pages} {033188} (\bibinfo {year}
  {2024})}\BibitemShut {NoStop}%
\bibitem [{\citenamefont {Vlasiuk}\ \emph {et~al.}(2023)\citenamefont
  {Vlasiuk}, \citenamefont {Kozin}, \citenamefont {Klinovaja}, \citenamefont
  {Loss}, \citenamefont {Iorsh},\ and\ \citenamefont
  {Tokatly}}]{PhysRevB.108.085410}%
  \BibitemOpen
  \bibfield  {author} {\bibinfo {author} {\bibfnamefont {E.}~\bibnamefont
  {Vlasiuk}}, \bibinfo {author} {\bibfnamefont {V.~K.}\ \bibnamefont {Kozin}},
  \bibinfo {author} {\bibfnamefont {J.}~\bibnamefont {Klinovaja}}, \bibinfo
  {author} {\bibfnamefont {D.}~\bibnamefont {Loss}}, \bibinfo {author}
  {\bibfnamefont {I.~V.}\ \bibnamefont {Iorsh}},\ and\ \bibinfo {author}
  {\bibfnamefont {I.~V.}\ \bibnamefont {Tokatly}},\ }\bibfield  {title}
  {\bibinfo {title} {Cavity-induced charge transfer in periodic systems:
  Length-gauge formalism},\ }\href
  {https://doi.org/10.1103/PhysRevB.108.085410} {\bibfield  {journal} {\bibinfo
   {journal} {Phys. Rev. B}\ }\textbf {\bibinfo {volume} {108}},\ \bibinfo
  {pages} {085410} (\bibinfo {year} {2023})}\BibitemShut {NoStop}%
\bibitem [{\citenamefont {Dmytruk}\ and\ \citenamefont
  {Schir{\`o}}(2022)}]{Dmytruk2022}%
  \BibitemOpen
  \bibfield  {author} {\bibinfo {author} {\bibfnamefont {O.}~\bibnamefont
  {Dmytruk}}\ and\ \bibinfo {author} {\bibfnamefont {M.}~\bibnamefont
  {Schir{\`o}}},\ }\bibfield  {title} {\bibinfo {title} {Controlling
  topological phases of matter with quantum light},\ }\href
  {https://doi.org/10.1038/s42005-022-01049-0} {\bibfield  {journal} {\bibinfo
  {journal} {Communications Physics}\ }\textbf {\bibinfo {volume} {5}},\
  \bibinfo {pages} {271} (\bibinfo {year} {2022})}\BibitemShut {NoStop}%
\bibitem [{\citenamefont {Todorov}\ and\ \citenamefont
  {Sirtori}(2014)}]{PhysRevX.4.041031}%
  \BibitemOpen
  \bibfield  {author} {\bibinfo {author} {\bibfnamefont {Y.}~\bibnamefont
  {Todorov}}\ and\ \bibinfo {author} {\bibfnamefont {C.}~\bibnamefont
  {Sirtori}},\ }\bibfield  {title} {\bibinfo {title} {Few-electron ultrastrong
  light-matter coupling in a quantum lc circuit},\ }\href
  {https://doi.org/10.1103/PhysRevX.4.041031} {\bibfield  {journal} {\bibinfo
  {journal} {Phys. Rev. X}\ }\textbf {\bibinfo {volume} {4}},\ \bibinfo {pages}
  {041031} (\bibinfo {year} {2014})}\BibitemShut {NoStop}%
\bibitem [{\citenamefont {Iqbal}\ \emph {et~al.}(2024)\citenamefont {Iqbal},
  \citenamefont {Todorov},\ and\ \citenamefont
  {Mora}}]{PhysRevResearch.6.033097}%
  \BibitemOpen
  \bibfield  {author} {\bibinfo {author} {\bibfnamefont {U.}~\bibnamefont
  {Iqbal}}, \bibinfo {author} {\bibfnamefont {Y.}~\bibnamefont {Todorov}},\
  and\ \bibinfo {author} {\bibfnamefont {C.}~\bibnamefont {Mora}},\ }\bibfield
  {title} {\bibinfo {title} {Dynamical coulomb blockade: An all electrical
  probe of the ultrastrong light-matter coupling regime},\ }\href
  {https://doi.org/10.1103/PhysRevResearch.6.033097} {\bibfield  {journal}
  {\bibinfo  {journal} {Phys. Rev. Res.}\ }\textbf {\bibinfo {volume} {6}},\
  \bibinfo {pages} {033097} (\bibinfo {year} {2024})}\BibitemShut {NoStop}%
\bibitem [{\citenamefont {Todorov}\ and\ \citenamefont
  {Sirtori}(2012)}]{PhysRevB.85.045304}%
  \BibitemOpen
  \bibfield  {author} {\bibinfo {author} {\bibfnamefont {Y.}~\bibnamefont
  {Todorov}}\ and\ \bibinfo {author} {\bibfnamefont {C.}~\bibnamefont
  {Sirtori}},\ }\bibfield  {title} {\bibinfo {title} {Intersubband polaritons
  in the electrical dipole gauge},\ }\href
  {https://doi.org/10.1103/PhysRevB.85.045304} {\bibfield  {journal} {\bibinfo
  {journal} {Phys. Rev. B}\ }\textbf {\bibinfo {volume} {85}},\ \bibinfo
  {pages} {045304} (\bibinfo {year} {2012})}\BibitemShut {NoStop}%
\bibitem [{\citenamefont {Todorov}\ \emph {et~al.}(2010)\citenamefont
  {Todorov}, \citenamefont {Andrews}, \citenamefont {Colombelli}, \citenamefont
  {De~Liberato}, \citenamefont {Ciuti}, \citenamefont {Klang}, \citenamefont
  {Strasser},\ and\ \citenamefont {Sirtori}}]{PhysRevLett.105.196402}%
  \BibitemOpen
  \bibfield  {author} {\bibinfo {author} {\bibfnamefont {Y.}~\bibnamefont
  {Todorov}}, \bibinfo {author} {\bibfnamefont {A.~M.}\ \bibnamefont
  {Andrews}}, \bibinfo {author} {\bibfnamefont {R.}~\bibnamefont {Colombelli}},
  \bibinfo {author} {\bibfnamefont {S.}~\bibnamefont {De~Liberato}}, \bibinfo
  {author} {\bibfnamefont {C.}~\bibnamefont {Ciuti}}, \bibinfo {author}
  {\bibfnamefont {P.}~\bibnamefont {Klang}}, \bibinfo {author} {\bibfnamefont
  {G.}~\bibnamefont {Strasser}},\ and\ \bibinfo {author} {\bibfnamefont
  {C.}~\bibnamefont {Sirtori}},\ }\bibfield  {title} {\bibinfo {title}
  {Ultrastrong light-matter coupling regime with polariton dots},\ }\href
  {https://doi.org/10.1103/PhysRevLett.105.196402} {\bibfield  {journal}
  {\bibinfo  {journal} {Phys. Rev. Lett.}\ }\textbf {\bibinfo {volume} {105}},\
  \bibinfo {pages} {196402} (\bibinfo {year} {2010})}\BibitemShut {NoStop}%
\bibitem [{\citenamefont {Todorov}\ \emph {et~al.}(2009)\citenamefont
  {Todorov}, \citenamefont {Andrews}, \citenamefont {Sagnes}, \citenamefont
  {Colombelli}, \citenamefont {Klang}, \citenamefont {Strasser},\ and\
  \citenamefont {Sirtori}}]{PhysRevLett.102.186402}%
  \BibitemOpen
  \bibfield  {author} {\bibinfo {author} {\bibfnamefont {Y.}~\bibnamefont
  {Todorov}}, \bibinfo {author} {\bibfnamefont {A.~M.}\ \bibnamefont
  {Andrews}}, \bibinfo {author} {\bibfnamefont {I.}~\bibnamefont {Sagnes}},
  \bibinfo {author} {\bibfnamefont {R.}~\bibnamefont {Colombelli}}, \bibinfo
  {author} {\bibfnamefont {P.}~\bibnamefont {Klang}}, \bibinfo {author}
  {\bibfnamefont {G.}~\bibnamefont {Strasser}},\ and\ \bibinfo {author}
  {\bibfnamefont {C.}~\bibnamefont {Sirtori}},\ }\bibfield  {title} {\bibinfo
  {title} {Strong light-matter coupling in subwavelength metal-dielectric
  microcavities at terahertz frequencies},\ }\href
  {https://doi.org/10.1103/PhysRevLett.102.186402} {\bibfield  {journal}
  {\bibinfo  {journal} {Phys. Rev. Lett.}\ }\textbf {\bibinfo {volume} {102}},\
  \bibinfo {pages} {186402} (\bibinfo {year} {2009})}\BibitemShut {NoStop}%
\bibitem [{\citenamefont {Todorov}(2015)}]{PhysRevB.91.125409}%
  \BibitemOpen
  \bibfield  {author} {\bibinfo {author} {\bibfnamefont {Y.}~\bibnamefont
  {Todorov}},\ }\bibfield  {title} {\bibinfo {title} {Dipolar quantum
  electrodynamics of the two-dimensional electron gas},\ }\href
  {https://doi.org/10.1103/PhysRevB.91.125409} {\bibfield  {journal} {\bibinfo
  {journal} {Phys. Rev. B}\ }\textbf {\bibinfo {volume} {91}},\ \bibinfo
  {pages} {125409} (\bibinfo {year} {2015})}\BibitemShut {NoStop}%
\bibitem [{\citenamefont {Goulain}\ \emph {et~al.}(2023)\citenamefont
  {Goulain}, \citenamefont {Deimert}, \citenamefont {Jeannin}, \citenamefont
  {Pirotta}, \citenamefont {Pasek}, \citenamefont {Wasilewski}, \citenamefont
  {Colombelli},\ and\ \citenamefont {Manceau}}]{Goulain2023}%
  \BibitemOpen
  \bibfield  {author} {\bibinfo {author} {\bibfnamefont {P.}~\bibnamefont
  {Goulain}}, \bibinfo {author} {\bibfnamefont {C.}~\bibnamefont {Deimert}},
  \bibinfo {author} {\bibfnamefont {M.}~\bibnamefont {Jeannin}}, \bibinfo
  {author} {\bibfnamefont {S.}~\bibnamefont {Pirotta}}, \bibinfo {author}
  {\bibfnamefont {W.~J.}\ \bibnamefont {Pasek}}, \bibinfo {author}
  {\bibfnamefont {Z.}~\bibnamefont {Wasilewski}}, \bibinfo {author}
  {\bibfnamefont {R.}~\bibnamefont {Colombelli}},\ and\ \bibinfo {author}
  {\bibfnamefont {J.}~\bibnamefont {Manceau}},\ }\bibfield  {title} {\bibinfo
  {title} {Thz ultra‐strong light–matter coupling up to 200 k with
  continuously‐graded parabolic quantum wells},\ }\bibfield  {journal}
  {\bibinfo  {journal} {Advanced Optical Materials}\ }\textbf {\bibinfo
  {volume} {11}},\ \href {https://doi.org/10.1002/adom.202202724}
  {10.1002/adom.202202724} (\bibinfo {year} {2023})\BibitemShut {NoStop}%
\bibitem [{\citenamefont {Andolina}\ \emph {et~al.}(2024)\citenamefont
  {Andolina}, \citenamefont {De~Pasquale}, \citenamefont {Pellegrino},
  \citenamefont {Torre}, \citenamefont {Koppens},\ and\ \citenamefont
  {Polini}}]{PoliniAndolinaAmperian2022}%
  \BibitemOpen
  \bibfield  {author} {\bibinfo {author} {\bibfnamefont {G.~M.}\ \bibnamefont
  {Andolina}}, \bibinfo {author} {\bibfnamefont {A.}~\bibnamefont
  {De~Pasquale}}, \bibinfo {author} {\bibfnamefont {F.~M.~D.}\ \bibnamefont
  {Pellegrino}}, \bibinfo {author} {\bibfnamefont {I.}~\bibnamefont {Torre}},
  \bibinfo {author} {\bibfnamefont {F.~H.~L.}\ \bibnamefont {Koppens}},\ and\
  \bibinfo {author} {\bibfnamefont {M.}~\bibnamefont {Polini}},\ }\bibfield
  {title} {\bibinfo {title} {Amperean superconductivity cannot be induced by
  deep subwavelength cavities in a two-dimensional material},\ }\href
  {https://doi.org/10.1103/PhysRevB.109.104513} {\bibfield  {journal} {\bibinfo
   {journal} {Phys. Rev. B}\ }\textbf {\bibinfo {volume} {109}},\ \bibinfo
  {pages} {104513} (\bibinfo {year} {2024})}\BibitemShut {NoStop}%
\bibitem [{\citenamefont {Jeannin}\ \emph {et~al.}(2019)\citenamefont
  {Jeannin}, \citenamefont {Mariotti~Nesurini}, \citenamefont {Suffit},
  \citenamefont {Gacemi}, \citenamefont {Vasanelli}, \citenamefont {Li},
  \citenamefont {Davies}, \citenamefont {Linfield}, \citenamefont {Sirtori},\
  and\ \citenamefont {Todorov}}]{TodorovExper2019}%
  \BibitemOpen
  \bibfield  {author} {\bibinfo {author} {\bibfnamefont {M.}~\bibnamefont
  {Jeannin}}, \bibinfo {author} {\bibfnamefont {G.}~\bibnamefont
  {Mariotti~Nesurini}}, \bibinfo {author} {\bibfnamefont {S.}~\bibnamefont
  {Suffit}}, \bibinfo {author} {\bibfnamefont {D.}~\bibnamefont {Gacemi}},
  \bibinfo {author} {\bibfnamefont {A.}~\bibnamefont {Vasanelli}}, \bibinfo
  {author} {\bibfnamefont {L.}~\bibnamefont {Li}}, \bibinfo {author}
  {\bibfnamefont {A.~G.}\ \bibnamefont {Davies}}, \bibinfo {author}
  {\bibfnamefont {E.}~\bibnamefont {Linfield}}, \bibinfo {author}
  {\bibfnamefont {C.}~\bibnamefont {Sirtori}},\ and\ \bibinfo {author}
  {\bibfnamefont {Y.}~\bibnamefont {Todorov}},\ }\bibfield  {title} {\bibinfo
  {title} {Ultrastrong light–matter coupling in deeply subwavelength thz lc
  resonators},\ }\href {https://doi.org/10.1021/acsphotonics.8b01778}
  {\bibfield  {journal} {\bibinfo  {journal} {ACS Photonics}\ }\textbf
  {\bibinfo {volume} {6}},\ \bibinfo {pages} {1207} (\bibinfo {year}
  {2019})}\BibitemShut {NoStop}%
\bibitem [{\citenamefont {Ashcroft}\ and\ \citenamefont
  {Mermin}(1976)}]{AshcroftMermin}%
  \BibitemOpen
  \bibfield  {author} {\bibinfo {author} {\bibfnamefont {N.}~\bibnamefont
  {Ashcroft}}\ and\ \bibinfo {author} {\bibfnamefont {N.}~\bibnamefont
  {Mermin}},\ }\href@noop {} {\emph {\bibinfo {title} {Solid State Physics}}}\
  (\bibinfo  {publisher} {Saunders Collage},\ \bibinfo {year}
  {1976})\BibitemShut {NoStop}%
\bibitem [{\citenamefont {Enkner}\ \emph {et~al.}(2024)\citenamefont {Enkner},
  \citenamefont {Graziotto}, \citenamefont {Boriçi}, \citenamefont
  {Appugliese}, \citenamefont {Reichl}, \citenamefont {Scalari}, \citenamefont
  {Regnault}, \citenamefont {Wegscheider}, \citenamefont {Ciuti},\ and\
  \citenamefont {Faist}}]{enkner2024enhancedfractionalquantumhall}%
  \BibitemOpen
  \bibfield  {author} {\bibinfo {author} {\bibfnamefont {J.}~\bibnamefont
  {Enkner}}, \bibinfo {author} {\bibfnamefont {L.}~\bibnamefont {Graziotto}},
  \bibinfo {author} {\bibfnamefont {D.}~\bibnamefont {Boriçi}}, \bibinfo
  {author} {\bibfnamefont {F.}~\bibnamefont {Appugliese}}, \bibinfo {author}
  {\bibfnamefont {C.}~\bibnamefont {Reichl}}, \bibinfo {author} {\bibfnamefont
  {G.}~\bibnamefont {Scalari}}, \bibinfo {author} {\bibfnamefont
  {N.}~\bibnamefont {Regnault}}, \bibinfo {author} {\bibfnamefont
  {W.}~\bibnamefont {Wegscheider}}, \bibinfo {author} {\bibfnamefont
  {C.}~\bibnamefont {Ciuti}},\ and\ \bibinfo {author} {\bibfnamefont
  {J.}~\bibnamefont {Faist}},\ }\href {https://arxiv.org/abs/2405.18362} {}
  (\bibinfo {year} {2024}),\ \Eprint {https://arxiv.org/abs/2405.18362}
  {arXiv:2405.18362} \BibitemShut {NoStop}%
\bibitem [{\citenamefont {Adam}\ \emph {et~al.}(2024)\citenamefont {Adam},
  \citenamefont {Duprez}, \citenamefont {Lehmann}, \citenamefont {Yglesias},
  \citenamefont {Cances}, \citenamefont {Ruckriegel}, \citenamefont
  {Masseroni}, \citenamefont {Tong}, \citenamefont {Denisov}, \citenamefont
  {Huang}, \citenamefont {Kealhofer}, \citenamefont {Garreis}, \citenamefont
  {Watanabe}, \citenamefont {Taniguchi}, \citenamefont {Ensslin},\ and\
  \citenamefont {Ihn}}]{adam2024entropyspectroscopybilayergraphene}%
  \BibitemOpen
  \bibfield  {author} {\bibinfo {author} {\bibfnamefont {C.}~\bibnamefont
  {Adam}}, \bibinfo {author} {\bibfnamefont {H.}~\bibnamefont {Duprez}},
  \bibinfo {author} {\bibfnamefont {N.}~\bibnamefont {Lehmann}}, \bibinfo
  {author} {\bibfnamefont {A.}~\bibnamefont {Yglesias}}, \bibinfo {author}
  {\bibfnamefont {S.}~\bibnamefont {Cances}}, \bibinfo {author} {\bibfnamefont
  {M.~J.}\ \bibnamefont {Ruckriegel}}, \bibinfo {author} {\bibfnamefont
  {M.}~\bibnamefont {Masseroni}}, \bibinfo {author} {\bibfnamefont
  {C.}~\bibnamefont {Tong}}, \bibinfo {author} {\bibfnamefont {A.~O.}\
  \bibnamefont {Denisov}}, \bibinfo {author} {\bibfnamefont {W.~W.}\
  \bibnamefont {Huang}}, \bibinfo {author} {\bibfnamefont {D.}~\bibnamefont
  {Kealhofer}}, \bibinfo {author} {\bibfnamefont {R.}~\bibnamefont {Garreis}},
  \bibinfo {author} {\bibfnamefont {K.}~\bibnamefont {Watanabe}}, \bibinfo
  {author} {\bibfnamefont {T.}~\bibnamefont {Taniguchi}}, \bibinfo {author}
  {\bibfnamefont {K.}~\bibnamefont {Ensslin}},\ and\ \bibinfo {author}
  {\bibfnamefont {T.}~\bibnamefont {Ihn}},\ }\href
  {https://arxiv.org/abs/2412.18000} {} (\bibinfo {year} {2024}),\ \Eprint
  {https://arxiv.org/abs/2412.18000} {arXiv:2412.18000} \BibitemShut {NoStop}%
\bibitem [{\citenamefont {Kittel}(2004)}]{kittel2004}%
  \BibitemOpen
  \bibfield  {author} {\bibinfo {author} {\bibfnamefont {C.}~\bibnamefont
  {Kittel}},\ }\href@noop {} {\emph {\bibinfo {title} {Introduction to Solid
  State Physics}}},\ \bibinfo {edition} {8th}\ ed.\ (\bibinfo  {publisher}
  {Wiley},\ \bibinfo {year} {2004})\BibitemShut {NoStop}%
\end{thebibliography}%

\end{document}